\newif\ifieee
\ieeefalse      
\documentclass[journal]{IEEEtran}
\ifieee
    \ifCLASSINFOpdf
    \else
    \fi
\else
\fi

\newif\ifreview
\reviewfalse 

\usepackage{amsmath}
\usepackage{cite}
\usepackage{amsfonts}
\usepackage{hyperref}
\hypersetup{
    colorlinks=true,
    linkcolor=blue,
    filecolor=magenta,      
    urlcolor=cyan,
    pdftitle={Overleaf Example},
    pdfpagemode=FullScreen,
    }

\usepackage{cleveref}

\usepackage[disable]{todonotes} 
\definecolor{pastel_mint_green}{RGB}{198, 240, 230} 
\definecolor{purple_custom}{RGB}{65,35,142}

\presetkeys{todonotes}{backgroundcolor=pastel_mint_green,linecolor=pastel_mint_green}{}

\newcounter{todocounter}

\let\oldtodo\todo  
\renewcommand{\todo}[1]{%
    \refstepcounter{todocounter}%
    \oldtodo[inline]{\thetodocounter \ - #1}%
}

\usepackage{amsthm}

\theoremstyle{definition}

\theoremstyle{remark}

\usepackage{mdframed}

\newmdtheoremenv[
  backgroundcolor=blue!8,
  linecolor=gray!60,
  linewidth=0.5pt,
  skipabove=8pt,
  skipbelow=8pt
]{boxedtheorem}{Theorem}

\newmdtheoremenv[
  backgroundcolor=blue!8,
  linecolor=gray!60,
  linewidth=0.5pt,
  skipabove=8pt,
  skipbelow=8pt
]{boxedproposition}{Proposition}

\newmdtheoremenv[
  backgroundcolor=green!8,
  linecolor=gray!60,
  linewidth=0.5pt,
  skipabove=8pt,
  skipbelow=8pt
]{boxedlemma}{Lemma}

\newmdtheoremenv[
  backgroundcolor=gray!8,
  linecolor=gray!60,
  linewidth=0.5pt,
  skipabove=8pt,
  skipbelow=8pt
]{boxeddefinition}{Definition}

\newmdtheoremenv[
  backgroundcolor=gray!8,
  linecolor=gray!60,
  linewidth=0.5pt,
  skipabove=8pt,
  skipbelow=8pt,
  nobreak=true,
]{boxedremark}{Remark}

\definecolor{Peach}{HTML}{FFDAB9}
\newmdtheoremenv[
  backgroundcolor=Peach!50,
  linecolor=gray!60,
  linewidth=0.5pt,
  skipabove=8pt,
  skipbelow=8pt
]{boxedcorollary}{Corollary}

\newmdtheoremenv[
  backgroundcolor=gray!8,
  linecolor=gray!60,
  linewidth=0.5pt,
  skipabove=8pt,
  skipbelow=8pt,
  innertopmargin=0.5\baselineskip,
  innerbottommargin=0.5\baselineskip
  ]{boxedassumption}{Assumption}

\newmdtheoremenv[
  backgroundcolor=gray!8,
  linecolor=gray!60,
  linewidth=0.5pt,
  skipabove=8pt,
  skipbelow=8pt
]{boxedsimplification}{Simplification}
\crefname{boxedtheorem}{Theorem}{Theorems}
\crefname{boxedproposition}{Proposition}{Propositions}
\crefname{boxedlemma}{Lemma}{Lemmas}
\crefname{boxedcorollary}{Corollary}{Corollary}
\crefname{boxeddefinition}{Definition}{Definitions}
\crefname{boxedremark}{Remark}{Remarks}
\crefname{boxedassumption}{Assumption}{Assumptions}
\crefname{boxedsimplification}{Simplification}{Simplifications}

\usepackage{tikz}
\usetikzlibrary{calc,matrix, positioning}

\usepackage{algorithm}
\usepackage{algpseudocode}

\usepackage{pgfplots}

\usepackage[utf8]{inputenc}
\usepackage[T1]{fontenc}

\usepackage{graphicx}
\usepackage{subcaption}

\newif\ifdesignexp
\designexpfalse 

\usepackage{booktabs}

\pgfplotsset{compat=1.18}

\usepackage{enumitem}

\usepackage{booktabs}
\usepackage{tabularx}

\begin{document}
%
\title{Reachability-Preserving Bellman Operator for the Discounted Reach-Cost Value Function: Uniting Hamilton–Jacobi Reachability and Reinforcement Learning}









%
%
%

\author{Isabelle~El-Hajj,
        Prashant~Solanki,
        Jasper van Beers,
        Coen de Visser,
        and~Erik-Jan~van Kampen
\thanks{All authors are with the section of Control \& Simulation at the Faculty of Aerospace Engineering at Delft University of Technology.}
}

%
%

\ifieee
    \markboth{IEEE Transactions on Automatic Control}%
    {El-Hajj \MakeLowercase{\textit{et al.}}: Unifying Hamilton--Jacobi Reachability and RL}
    %
\fi



\maketitle


\begin{abstract}
Hamilton-Jacobi (HJ) reachability provides rigorous safety and reachability guarantees for continuous-time dynamical systems, but its numerical solution suffers from the curse of dimensionality. Deep reinforcement learning (DRL), by contrast, offers scalable sample-based methods. However, RL is typically built around additive cumulative rewards; whereas, reachability objectives are inherently non-additive. This mismatch makes a direct bridge between HJ reachability and RL nontrivial. Recent discounted formulations have either introduced contraction by altering the original reachability semantics, or preserved exact semantics on the HJ side without a corresponding Bellman fixed-point characterization. In this paper, we close this gap by building on a semantics-preserving discounted reach-based value function and deriving a non-additive Bellman operator whose unique fixed point exactly matches the value function in the HJ formulation. We prove that discounting makes this operator contractive, yielding existence, uniqueness, and convergence of value iteration. Furthermore, we establish the equivalence between the HJ and Bellman characterizations, and show that RL can be interpreted as a sample-based approximation scheme for the same fixed-point equation. This yields a principled and semantically exact connection between HJ reachability and RL, enabling learning-based methods to approximate reachability value functions while preserving their safety-critical meaning. As a result, the proposed framework opens the door to scalable, data-driven computation of reachable sets and safety certificates in high-dimensional systems. Numerical experiments demonstrate close agreement with Hamilton–Jacobi solutions, confirm preservation of reachability semantics via alignment of zero level sets, and support the interpretation of reinforcement learning as a sample-based solver of the proposed Bellman operator.
\end{abstract}

\begin{IEEEkeywords}
Hamilton-Jacobi reachability, Non-additive Bellman Operator, Reinforcement learning, Reach cost, Safety-critical control.
\end{IEEEkeywords}

\section{Introduction}
\label{sec:introduction}

\subsection{Background \& Motivation}

Hamilton--Jacobi (HJ) reachability provides a principled framework for analyzing safety and reachability properties of continuous-time dynamical systems \cite{lygeros2004reachability}. By characterizing reachable or safe sets as sublevel or superlevel sets of a value function satisfying a Hamilton--Jacobi variational inequality (HJVI), this framework offers strong semantic guarantees and has become a cornerstone of safety-critical control \cite{bansal2017hamilton,choi2021robust}. However, its practical utility remains limited by the curse of dimensionality: numerical solutions of the associated partial differential equations scale exponentially with the state dimension \cite{ganai2024hamilton,ganai2024hamiltonPhD}.

A large body of work has therefore sought to improve the scalability of HJ reachability, including decomposition methods, level-set approximations, and learning-based surrogates \cite{lee2020hopf,herbert2021scalable,xue2019inner,li2026converse,darbon2016algorithms}. Learning methods that are constrained by partial differential equation (PDEs), such as DeepReach \cite{bansal2021deepreach}, approximate the time-dependent HJ value function using a neural implicit representation trained to satisfy the PDE together with terminal and boundary conditions. While such approaches enforce local PDE consistency, they do not by themselves provide a global Bellman fixed-point characterization of the solution.

In parallel, deep reinforcement learning (DRL) has emerged as a powerful paradigm for sequential decision-making and control in high-dimensional systems. Across games, robotics, fluid dynamics, quantum control, and industrial processes, DRL has shown that it can address problems that are difficult for classical control and optimization \cite{Mnih2015Human-level,degrave2022magnetic,silver2016mastering,silver2017mastering,silver2018general,vinyals2019grandmaster,andrychowicz2020learning}. By leveraging function approximation and sample-based updates, DRL methods can scale to settings that are often intractable for classical dynamic programming. However, standard RL formulations are typically built around additive cumulative rewards; whereas, reachability objectives are fundamentally non-additive: they depend on whether a target can be reached or avoided over a time horizon, rather than on the accumulation of instantaneous rewards alone. As a result, a direct bridge between HJ reachability and RL is not immediate. Any such bridge must preserve the semantics of reachability while also admitting an operator-theoretic formulation compatible with dynamic programming and learning.

\subsection{Related Work \& Contributions}

The complementary strengths and limitations of HJ reachability and RL have motivated considerable interest in bridging the two frameworks \cite{fisac2019bridging,ganai2024hamilton,solanki2026formalizing,sharpless2025dual,sharpless2026bellman}. Broadly speaking, prior work has shown either that discounting can induce contraction properties and RL-compatible Bellman operators, or that discounted reachability formulations can preserve the exact reachability semantics on the HJ side. What has remained unaddressed is an exact Bellman fixed-point characterization for a semantics-preserving discounted reach-based value function.

Nonetheless, several works have taken important steps in this direction. Akametalu et al.\ introduced the minimum discounted reward (MDR) formulation, which recasts reachability as a discounted reward optimization problem and yields a contractive Bellman operator amenable to standard reinforcement learning methods \cite{akametalu2024minimum}. Crucially, this work demonstrates that discounting can induce a contraction structure. However, the resulting value function does not exactly recover the original reachability semantics, and instead provides a discounted approximation of the underlying reachability problem.

Fisac et al.\ \cite{fisac2019bridging} further highlighted the interface between time-discounted safety analysis and RL by showing how discounted formulations can make RL tools amenable to approximating safety value functions. Their work provides an important conceptual bridge between HJ safety analysis and RL, and helped motivate subsequent work at this intersection. In discrete time, Hsu et al.\ \cite{hsu2021safety} derived discounted Bellman equations for reach-avoid problems and showed how such objectives can be approached through RL. More recently, related operator-theoretic perspectives have also appeared in Bellman-style formulations for constrained and reachability-related sequential decision problems \cite{sharpless2025dual,sharpless2026bellman}. Collectively, these works demonstrate that reachability-type objectives can be incorporated into Bellman-style learning frameworks, but they do not provide the exact continuous-time HJ/Bellman equivalence established here for the discounted reach-based setting.

On the semantics-preserving HJ side, Choi et al.\ \cite{choi2021robust} introduced a discounted reachability-based value function, termed the Control Barrier-Value Function (CBVF), and showed that it satisfies a Hamilton--Jacobi--Isaacs variational inequality (HJIVI). Importantly, this formulation preserves reachability semantics under exponential discounting and establishes a connection between HJ reachability and control barrier functions. Thus, Choi et al.\ provide the appropriate discounted HJ formulation upon which an exact bridge to RL can be built. 

Prior work achieves either contraction or semantic correctness, but not both simultaneously within a Bellman fixed-point framework. We take the semantics-preserving discounted HJ characterization formulation from Choi et al. \cite{choi2021robust} as our starting point and provide the missing Bellman fixed-point characterization.

This work is also complementary to recent efforts that formalize the relationship between HJ reachability and RL through a travel-cost (also referred to as running-cost) formulation \cite{solanki2026formalizing}. While that line of work focuses on a travel-cost-based value function, we instead consider a reach-cost-based value function. Both viewpoints preserve reachability semantics and admit contraction-based operator structures compatible with modern learning methods, but they lead to different value-function definitions and different Bellman operators.

In this paper, we build on the semantics-preserving discounted reach-based formulation of Choi et al.\ and provide the missing operator-theoretic bridge to reinforcement learning. Specifically, we derive a reachability-preserving Bellman operator whose fixed point exactly characterizes the same discounted reach-based value function that appears in the HJ formulation. Although the operator is non-additive, discounting renders it contractive, enabling the use of Banach's fixed-point theorem \cite{granas2003fixed} to establish existence, uniqueness, and convergence of value iteration. This yields a direct and principled bridge between HJ reachability and reinforcement learning.

The main contributions of this paper are as follows:
\begin{itemize}
    \item Starting from the discounted reach-based value function and HJ characterization introduced by Choi et al.\ \cite{choi2021robust}, we derive a reachability-preserving Bellman operator corresponding exactly to it.
    \item We prove that, although this Bellman operator is non-additive, discounting renders it a contraction on a complete space of value functions, yielding a unique fixed point via Banach's fixed-point theorem.
    \item We establish the equivalence between the HJ characterization and the Bellman fixed-point characterization of this discounted reach-based value function.
    \item We show that reinforcement learning algorithms can be interpreted as sample-based approximation schemes for this Bellman operator, thereby yielding a principled and semantically correct bridge between HJ reachability and reinforcement learning.
\end{itemize}

The remainder of the paper is structured as follows. \Cref{sec:problem-statement} introduces the problem formulation and elucidates the notation used to distinguish different successive forms of the value function or candidates thereof, including the finite- and infinite-horizon variants developed in parallel throughout the paper. \Cref{sec:reach-based-value-function} defines the discounted reach-based value function, establishes its key properties, and derives the associated dynamic programming principle (DPP). In \Cref{sec:hj-characterization}, the associated HJ characterization is presented. \Cref{sec:bellman-exact} then introduces the corresponding Bellman operator and proves the existence and uniqueness of its fixed point. The equivalence between the HJ and Bellman formulations is established in \Cref{sec:equivalence}. \Cref{sec:bellman-numerical} further develops a discrete-time version of the Bellman operator, shows its convergence to the exact operator. \Cref{sec:rl-sampled-based} discusses learning the fixed point via reinforcement learning through sample-based access. \Cref{sec:exp-res} presents numerical validation of the proposed theoretical framework. Finally, \Cref{sec:conclusion} offers concluding remarks and directions for future work.

\section{Problem Formulation \& Notation}
\label{sec:problem-statement}

\subsection{System Dynamics}
We consider a controlled dynamical system of the form
\begin{equation}
    \label{eq:dynamics}
    \dot{x}(s) = f(x(s),u(s)), \qquad s \in [t, T],
\end{equation}
where $x(s) \in \mathbb{R}^n$ is the state trajectory and $u(s) \in \mathcal{U} \subset \mathbb{R}^m$ is the control input. We define $\mathcal{M}(t)$ as the set of all control policies applicable at time $t$:
\[
    \mathcal{M}(t)\equiv\{u:[t,T]\rightarrow \mathcal{U}\mid u \text{ measurable} \}.
\]

We make the following standing assumptions on the vector field $f$, beginning with its functional form.

\begin{boxedassumption}[Time-invariance of $f$]
\label{ass:f-time-invariant}
The vector field $f$ is a function of state and control alone,
\[
f:\mathbb R^n\times\mathcal U \to \mathbb R^n,
\]
with no explicit dependence on time, as reflected in \cref{eq:dynamics}.
\end{boxedassumption}

\begin{boxedassumption}[Global Lipschitz continuity of $f$ in $x$]
\label{ass:f-lipschitz-x}
The function $f : \mathbb{R}^{n} \times \mathcal{U} \to \mathbb{R}^n$ is globally Lipschitz continuous in $x$, uniformly in $u$. That is, there exists a constant $L_{fx} > 0$ such that
\begin{equation}
\begin{aligned}
    \|f(x,u) - f(y,u)\| &\le L_{fx} \|x - y\| \\
    &\quad \forall x,y \in \mathbb{R}^{n},\; \forall u \in \mathcal{U}.
\end{aligned}
\end{equation}
\end{boxedassumption}

\begin{boxedassumption}[Continuity of $f$ in $u$]
\label{ass:f-cont-u}
For each $x \in \mathbb{R}^{n}$, the function $f(x,u)$ is continuous in $u$.
\end{boxedassumption}

\begin{boxedassumption}[Compactness of $\mathcal{U}$]\label{ass:control-set-compact}
The control set $\mathcal{U} \subset \mathbb{R}^{m}$ is nonempty and compact (i.e. closed and bounded).
\end{boxedassumption}

\subsection{Target Set Representation}
We represent the target set $\mathcal{G} \subset \mathbb{R}^{n}$ through a reach cost function $g$ such that
\[
\mathcal G=\{x\in\mathbb R^n\mid g(x)<0\}.
\]
The sign of $g$ therefore determines whether a state lies inside or outside the target set.

\begin{boxedassumption}[Sign Calibration / Semantic Assumption]\label{ass:g-sign-calibration}
\leavevmode
\begin{itemize}[topsep=0pt, itemsep=0pt, parsep=0pt, partopsep=0pt, before=\vspace{-\baselineskip}]
    \item $g(x) < 0$ inside target set $\mathcal{G}$
    \item $g(x) \geq 0$ outside target set $\mathcal{G}$
\end{itemize}
\end{boxedassumption}

We make the following additional assumptions on $g$, beginning with its functional form.

\begin{boxedassumption}[Time-independence of $g$]
\label{ass:g-time-independent}
The reach cost is a function of state alone,
\[
g:\mathbb R^n \to \mathbb R,
\]
with no explicit dependence on time; equivalently, the target set $\mathcal G$ is fixed for all $t$.
\end{boxedassumption}

\begin{boxedassumption}[Global Lipschitz continuity of $g$ in $x$]
\label{ass:g-lipschitz-x}
The reach cost $g(x) : \mathbb{R}^{n} \to \mathbb{R}$ is Lipschitz continuous in $x$. That is, there exists a constant $L_g > 0$ such that
\begin{equation}
\|g(x) - g(y)\| \le L_{g} \|x - y\| \quad \forall x,y \in \mathbb{R}^{n}.
\end{equation}
\end{boxedassumption}

\begin{boxedassumption}[Boundedness of $g$]
\label{ass:g-bounded}
    $g$ is bounded, i.e.
    \begin{equation}
       \|g\|_\infty := \sup_{x\in\mathcal X}|g(x)| \leq M < \infty.
    \end{equation}
    for some finite $M$.
\end{boxedassumption}

\subsection{Reachability Objective}
\label{subsec:reachability-objective}

Let $T \in [0,\infty]$ denote the terminal time and $\xi_{t,x}^{u}(s)$ denote the trajectory of the system under the admissible control
$u(\cdot)$ starting from state $x$ at time $t$, evaluated at absolute time $s$. The dynamics can be written as follows:
\[
\frac{d}{ds}\xi_{t,x}^{u}(s)=f\bigl(\xi_{t,x}^{u}(s),u(s)\bigr),
\qquad \xi_{t,x}^{u}(t)=x,
\]
for all $s \in [t,T]$.

We consider the reachability problem over the horizon $[t,T]$.

\noindent\textbf{Backward reachable tube (BRT).} The BRT is the set of states from which there exists an admissible control that drives the system into the target set $\mathcal{G}$ at some time within the horizon:
\begin{equation}
    \begin{aligned}
        &\mathcal{R}_{\mathrm{BRT}}(t)
        := \\
        &\left\{
        x \in \mathbb{R}^{n}
        \;\middle|\;
        \exists u(\cdot),\ \exists \tau \in [t,T]
        \text{ such that }
        \xi_{t,x}^{u}(\tau)\in \mathcal{G}
        \right\}.
    \end{aligned}
\end{equation}

\begin{boxedremark}
For clarity of presentation, the theory and all main results in this paper are formulated and developed for the backward reachable tube (BRT) problem. The corresponding avoid formulation is entirely analogous. In particular, the same arguments and results carry over after replacing the reach objective by the safety objective, the target set by the unsafe set where appropriate, and the control optimization $\inf$ by $\sup$ in the relevant definitions and results.
\end{boxedremark}

\subsection{Notation Across the Value Function's Successive Forms}
Having introduced the system dynamics, target set, and reachability objective, we close this section with a summary of the notation used for the value function and its successive forms, developed in \cref{sec:reach-based-value-function,sec:hj-characterization,sec:bellman-exact,sec:equivalence,sec:bellman-numerical}. The value function is not a single fixed object throughout the paper: it appears first in absolute time, then in a time-to-go reparameterization, then as two \emph{a priori} distinct
characterizations---a viscosity solution and a Bellman fixed point---that are subsequently shown to coincide, and finally as a discretized and a learned approximation thereof. The notation in \cref{tab:notation-roadmap} tracks this progression directly, with each tier of symbols corresponding to a distinct level of exactness; the table is intended as a running reference as new tiers are introduced across those sections.

\begin{table}[t]
    \centering
    \footnotesize
    \caption{Value-Function Notation Across Successive Forms.}
    \label{tab:notation-roadmap}
    \renewcommand{\arraystretch}{1.2}

    \begin{tabular}{@{}c l >{\raggedright\arraybackslash}p{0.55\columnwidth}@{}}
    \toprule
    Tier & Symbol & Meaning \\
    \midrule

    0 &
    $V(t,x)$, $V^{\infty}(t,x)$ &
    True value function expressed in absolute time (\cref{def:V,def:V-infinity}) \\
    
    1 &
    $W(\tau,x)$, $W^\infty(x)$ &
    True value function expressed in the time-to-go variable $\tau$ (\cref{def:W}) and the true stationary time-to-go value function (\cref{def:W-inf}) \\
    

    2 &
    \begin{tabular}[t]{@{}l@{}}
    $W_{\mathrm{Bell}}(\tau,x)$, \\
    $W_{\mathrm{Bell}}^{\infty}(x)$
    \end{tabular}
    &
    The unique Bellman fixed point (\cref{thm:unique-fixed-point-T}, \cref{thm:unique-fixed-point-T-infinity}),
    a priori distinct from $W(\tau,x)$ and $W^\infty(x)$ until shown to coincide with them
    in \cref{thm:equivalence-characterization} and \cref{thm:equivalence-characterization-infinity}
    respectively.\\
    
    3 &
    $\widehat{W}(\tau,x)$, $\widehat{W}^{\infty}(x)$ &
    Fixed point of the finite-horizon discrete Bellman operator $\widehat{T}_{\sigma,\lambda}$ (\cref{eq:W-hat}) and of the infinite-horizon discrete Bellman operator $\widehat{T}^{\infty}_{\sigma,\lambda}$ \\
    
    4 &
    $\Psi_\theta(\tau, x)$ , $\Psi^{\infty}_\theta(x)$ &
    Parametric (learned) approximation produced by fitted-value iteration (FVI) for finite horizon and infinite horizon cases (\cref{sec:rl-sampled-based}) \\
    
    --- &
    $\Psi$ &
    Generic placeholder argument to a Bellman operator \\
    
    \bottomrule
    \end{tabular}
\end{table}

\section{Discounted Reach-based Value Function}
\label{sec:reach-based-value-function}

The set-based reachability problem discussed in \cref{subsec:reachability-objective} admits an equivalent value-function characterization which we present in this section (see \cref{def:V,def:V-infinity}). In particular, the zero sublevel set of this equivalent value function coincides with the backward reachable tube \cite{herbert2021scalable,bansal2017hamilton,lygeros2004reachability,choi2021robust}:
\begin{equation*}
    \mathcal{R}_{\mathrm{BRT}}(t)=\{x\in\mathcal{X}\mid V(t,x)\le 0\}.
\end{equation*}
Thus, the sign of the value function determines membership in the reachable set: $V(t,x)\le 0$ if and only if the state $x$ belongs to the backward reachable tube at time $t$.

\subsection{Absolute-Time Value Function}

\begin{boxeddefinition}[Discounted Reach-based Value Function]
\label{def:V}
\begin{equation}
    V(t, x) := \inf_{u(\cdot) \in \mathcal{U}[t, T]} \inf_{s \in [t, T]} g(\xi_{t, x}^{u}(s)) e^{-\lambda (s-t)},
    \label{eq:value_function}
\end{equation}
where $\lambda > 0$, and we refer to is as the \textit{discount rate}. 
\end{boxeddefinition}

\noindent\textit{Interpretation.}
The value function evaluates the most favorable (i.e., smallest) discounted reach cost attainable along a trajectory. The inner infimum over time introduces a stopping structure, capturing the first favorable encounter with the target, while discounting ensures well-posedness and contraction without altering the sign-based reachability semantics.

\begin{boxeddefinition}[Infinite-Horizon Value Function $V^{\infty}(t,x)$]
\label{def:V-infinity}
The infinite-horizon value function is \cref{def:V} evaluated in the limit $T \to \infty$:
\begin{equation}
    V^{\infty}(t,x) := \inf_{u(\cdot)\in\mathcal{U}[t,\infty)} \inf_{s\in[t,\infty)} g(\xi_{t,x}^u(s))\, e^{-\lambda(s-t)}.
    \label{eq:V_infinity}
\end{equation}
$V^{\infty}$ formally retains $t$ as an argument, consistent with \cref{def:V}. However, \cref{lem:V-infinity-t-invariant} shows it is in fact invariant of $t$.
\end{boxeddefinition}

\subsection{Time-to-Go Value Function}

Since reinforcement learning takes on a time-to-go perspective, we introduce the \emph{remaining time variable} $\tau:=T-t$ where $t$, in consistency with the previous semantics, denotes the absolute time. We substitute $t=T-\tau$ into \cref{eq:value_function}.

\begin{equation}
\begin{aligned}
    V(T-\tau, x) &= \inf_{u(\cdot) \in \mathcal{U}} \inf_{s \in [T-\tau, T]} g(\xi_{T-\tau, x}^{u}(s)) e^{\lambda (T-\tau-s)}
\end{aligned}
\label{eq:value_function_T_min_tau}
\end{equation}
\Cref{eq:value_function_T_min_tau} is a direct re-expression of the original reachability value function now in terms of remaining time, yet without altering its time semantics: the first argument of $V(\cdot, \cdot)$ continues to represent absolute time, and the trajectory $\xi_{T-\tau, x}^{u}(s)$ remains parameterized by absolute time.

Next, we introduce the \emph{elapsed-time variable $r$} which is given by
\begin{equation}
    r:=s-(T-\tau) \qquad \text{where} \ s\in[t, T] 
\end{equation}

This then maps the interval $s\in[T-\tau, T]$ to $r \in [0, \tau]$, which leads to the following updated version of \cref{eq:value_function_T_min_tau}.

\begin{equation}
\begin{aligned}
    V(T-\tau, x) &= \inf_{u(\cdot) \in \mathcal{U}} \inf_{r \in [0, \tau]} g(\xi_{T-\tau, x}^{u}(T - \tau + r)) e^{-\lambda r} .
\end{aligned}
\label{eq:value_function_T_min_tau_with_r}
\end{equation}

Since the dynamics are time-invariant (\cref{ass:f-time-invariant}), trajectories depend only on the initial state, the elapsed time, and the control function. For any admissible control $u(\cdot)$, we therefore define the flow
\begin{equation}
    \phi_x^u(r) := \xi_{T-\tau, x}^{u}(T-\tau + r),
\end{equation}
which represents the state reached after elapsed time $r$ starting from $x$.

With this notation, and since $g$ is time-independent (\cref{ass:g-time-independent}), we can rewrite
\begin{equation}
    g\left(\xi_{T-\tau, x}^u(T-\tau+r)\right)=g\left(\phi_x^u(r)\right),
\end{equation}
We can now define the time-to-go value function $W(\tau,x)$.

\begin{boxeddefinition}[Finite-Horizon Time-to-Go Value Function $W(\tau, x)$]
\label{def:W}
\begin{equation}
        W(\tau, x) :=V(T-\tau, x) = \inf_{u(\cdot) \in \mathcal{U}[0,\tau]} \inf_{r \in [0, \tau]} g(\phi_{x}^{u}(r)) e^{-\lambda r} .
        \label{eq:W_eqn}
\end{equation}
\end{boxeddefinition}
\noindent\textit{Interpretation.} In words, $W(\tau, x)$ is the smallest discounted reach-cost value that can be attained within the next $\tau$ units of time, starting from state $x$, under any admissible control.

\begin{boxeddefinition}[Infinite-Horizon Time-to-Go Value Function $W^{\infty}(x)$]
\label{def:W-inf}
For the infinite-horizon case, the time-to-go variable $\tau$ becomes unbounded; we write $W^{\infty}(x)$ for \cref{def:W} evaluated at $\tau=\infty$, which becomes only a function of $x$
\begin{equation}
    W^{\infty}(x) := \inf_{u(\cdot)\in\mathcal{U}} \inf_{r\in[0,\infty)} g(\phi_x^u(r))\, e^{-\lambda r}.
    \label{eq:W_infinity}
\end{equation}
\end{boxeddefinition}

\subsection{Properties of the Value Function}
In this subsection, we establish several regularity properties of the value function $V(t,x)$ and its infinite-horizon counterpart $V^{\infty}(t,x)$, developed in parallel for the finite- and infinite-horizon settings. For finite horizon $T<\infty$, we establish boundedness (\cref{lem:V-bounded}) and Lipschitz continuity, both in the state variable (\cref{lem:V-lipschitz-continuous-x-finite-T}) and in time, the latter stated jointly for $V(t,x)$ and the time-to-go value function $W(\tau,x)$ (\cref{lem:V-lipschitz-continuous-t-infinite-T}). For the infinite-horizon case, we establish time-invariance of $V^{\infty}(t,x)$ in $t$ (\cref{lem:V-infinity-t-invariant}) — which underlies its reduction to the stationary function $W^{\infty}(x)$ — together with pointwise continuity in $x$ (\cref{lem:V-continuous-x-infinite}) and in $t$ (\cref{lem:V-continuous-t-infinite}), the latter following directly from time-invariance. 

\begin{boxedlemma}[Boundedness of $V(t,x)$]
\label{lem:V-bounded}
Assume that the reach cost $g$ is bounded (\cref{ass:g-bounded}). Then, the value function defined in \cref{eq:value_function} is bounded, i.e. 
\begin{equation}
    \|V\|_{\infty} \leq M < \infty 
\end{equation}
\end{boxedlemma}

\begin{proof}
Since $s - t \geq 0$ and $\lambda >0$, then $e^{-\lambda(s-t)}  \leq 1$. Therefore, for all admissible control $u(\cdot) \in \mathcal{U}$ and $s \in [t, T]$
\begin{equation}
   | g(\xi_{t, x}^{u}(s)) e^{-\lambda (s-t)}| \leq |g(\xi_{t, x}^{u}(s)) | \leq M
\end{equation}
Taking the infimum over $u(\cdot) \in \mathcal{U}$ and $s \in [t, T]$ preserves this bound:
    \begin{equation}
        |V(t,x)| \leq M \qquad \forall (t,x)
    \end{equation}
Taking the supremum over $(t, x)$ gives
\begin{equation}
    \|V \|_{\infty} \leq M < \infty
\end{equation}
\end{proof}

\begin{boxedlemma}[Time-invariance of $V^{\infty}(t,x)$]
    \label{lem:V-infinity-t-invariant}
\end{boxedlemma}
\begin{proof}
    The expression for $V^{\infty}(t,x)$ can be seen in \cref{def:V-infinity}. We set \(r=s-t\). Since the dynamics are time-invariant, admissible controls can be shifted in
    time. Thus, with \(\bar u(r):=u(t+r)\),
    \[
    \xi^u_{t,x}(t+r)=\phi_x^{\bar u}(r).
    \]
    Therefore,
    \[
    \begin{aligned}
    V^{\infty}(t,x)
    &=
    \inf_{\bar u(\cdot)}
    \inf_{r\in[0,\infty)}
    e^{-\lambda r}g(\phi_x^{\bar u}(r)) \\
    &=:W^{\infty}(x).
    \end{aligned}
    \]

    Hence, \(V^{\infty}(t,x)\) is invariant with respect to $t$, and $W^{\infty}(x)$ is independent of $\tau$.
\end{proof}

\begin{boxedlemma}[Lipschitz Continuity of $V(t,x)$ in $x$ for finite horizon $T$]
\label{lem:V-lipschitz-continuous-x-finite-T}
The value function $V(t,x)$ defined in \cref{eq:value_function} is globally Lipschitz continuous in $x$ for each fixed $t \in [0,T]$ for $T < \infty$. That is, there exists $L_{Vx} > 0$ such that
\begin{equation}
\begin{aligned}
    \|V(t, x) - V(t, y)\| &\le L_{Vx} \|x - y\|, \\
    &\forall x,y \in \mathbb{R}^{n}, \quad t \in [0, T] \text{ s.t. } T < \infty
\end{aligned}
\end{equation}
\end{boxedlemma}

\begin{proof}
    Using the Lipschitz continuity of the dynamics in $x$ (\cref{ass:f-lipschitz-x}) and Gr\"{o}nwall's inequality, it is possible to establish a trajectory difference bound

    \begin{equation}
        \label{eq:traj-diff-bound}
        \left\|\xi_{t, x}^u(s)-\xi_{t, y}^u(s)\right\| \leq e^{L_{fx}(s-t)}\|x-y\|
    \end{equation}
    Applying the Lipschitz continuity of $g$ in $x$ (\cref{ass:g-lipschitz-x}) and substituting \cref{eq:traj-diff-bound} yields
    \begin{equation}
        \label{eq:bound-on-g-diff}
        \left|g\left(\xi_{t, x}^u(s)\right)-g\left(\xi_{t, y}^u(s)\right)\right| \leq L_g e^{L_{fx}(s-t)}\|x-y\|
    \end{equation}
    Incorporating the discount factor
    \begin{equation}
        \begin{gathered}
        \left|g\left(\xi_{t, x}^u(s)\right) e^{-\lambda(s-t)}-g\left(\xi_{t, y}^u(s)\right) e^{-\lambda(s-t)}\right| \\
        \leq L_g e^{\left(L_{f x}-\lambda\right)(s-t)}\|x-y\|
        \end{gathered}
    \end{equation}
    Applying the property that 
    \begin{equation}
        |\inf A-\inf B| \leq \sup _{a \in A, b \in B}|a-b|
    \end{equation}
    this means that
    \begin{equation}
        |V(t, x)-V(t, y)| \leq \sup _u \sup _{s \in[t, T]} L_g e^{\left(L_{f x}-\lambda\right)(s-t)}\|x-y\|
    \end{equation}

    Since the right-hand-side is independent of $u$, this above simplifies to
    \begin{equation}
        |V(t, x)-V(t, y)| \leq \sup _{s \in[t, T]} L_g e^{\left(L_{f x}-\lambda\right)(s-t)}\|x-y\|
    \end{equation}

    The result for $\sup _{s \in[t, T]} L_g e^{\left(L_{f x}-\lambda\right)(s-t)}$ will depend on how $L_{fx}$ and $\lambda$ compare. Specifically,

    \begin{equation}
        \label{eq:cases_with_t}
        \sup _{s \in[t, T]} L_g e^{\left(L_{f x}-\lambda\right)(s-t)} =
        \begin{cases}
            L_g, & \lambda \ge L_{fx}, \\
            L_g e^{(L_{fx}-\lambda)(T-t)}, & \lambda < L_{fx}.
        \end{cases}    
    \end{equation}
    We can upper-bound the second case:
    \begin{equation}
        \label{eq:cases_no_t}
        \sup _{s \in[t, T]} L_g e^{\left(L_{f x}-\lambda\right)(s-t)} \leq
        \begin{cases}
            L_g, & \lambda \ge L_{fx}, \\
            L_g e^{(L_{fx}-\lambda)T}, & \lambda < L_{fx}.
        \end{cases}    
    \end{equation}
    It is clear that for finite $T$, that $V(t,x)$ is Lipschitz continuous in $x$ uniformly in time and hence is continuous in $x$.
\end{proof}

\begin{boxedremark}[Lipschitz continuity of $V^{\infty}(t,x)$ in $x$]
    As for the case of when $T \to \infty$, $e^{(L_{f x}-\lambda)(T-t)}$ would be finite only if $\lambda \geq L_{fx}$. Hence, in the infinite-horizon setting, the discount rate must dominate the growth induced by the dynamics in order to guarantee a global Lipschitz bound for the value function.

    This regularity, however, is too strong and unnecessary for viscosity theory. Instead, we show the pointwise continuity of $V(t,x)$ in $x$ for the case of $T \to \infty$.
\end{boxedremark}

\begin{boxedlemma}[Pointwise Continuity of $V^{\infty}(t,x)$ in $x$]
\label{lem:V-continuous-x-infinite}
For every fixed $t \geq 0$, the value function $V(t,x)$ (\cref{def:V}) is, in the limit $T \to \infty$, pointwise continuous in $x$ on $\mathbb{R}^n$.
\end{boxedlemma}

\begin{proof}

Fix $t \geq 0$ and fix a state $x \in \mathbb{R}^n$. We want to show: for every $\varepsilon > 0$, there exists $\delta > 0$ such that
\[
\|x-y\| < \delta
\;\Longrightarrow\;
|V(t,x)-V(t,y)| < \varepsilon.
\]
Define the truncated finite-horizon value function

\[
V_S(t,x)
:=
\inf_{u(\cdot)\in\mathcal U[t,S]}
\inf_{s\in[t,S]}
g(\xi^u_{t,x}(s))e^{-\lambda(s-t)},
\]
for some finite $S > t$.

Let $\varepsilon > 0$ be given. By the triangle inequality:
\begin{equation}
    \begin{aligned}
        & |V(t,x)-V(t,y)| \leq \underbrace{|V(t,x)-V_S(t,x)|}_{(1)} \\
        &+ \underbrace{|V_S(t,x)-V_S(t,y)|}_{(2)} + \underbrace{|V_S(t,y)-V(t,y)|}_{(3)}.
    \end{aligned}
\end{equation}

\textbf{Controlling terms (1) and (3).}
Since $g$ is bounded by $M$ (\cref{ass:g-bounded}) and the discount factor satisfies
\[
e^{-\lambda(s-t)} \leq e^{-\lambda(S-t)}
\quad \text{for all } s \geq S,
\]
the tail contribution beyond $S$ satisfies:
\[
|V(t,x)-V_S(t,x)|
\leq
M e^{-\lambda(S-t)}.
\]

The same bound holds for term $(3)$ with $y$ in place of $x$, since the bound is independent of the state. Choose $S$ large enough so that
\[
M e^{-\lambda(S-t)} < \frac{\varepsilon}{3},
\]
which is possible since
\[
M e^{-\lambda(S-t)} \to 0
\quad \text{as } S \to \infty.
\]
This choice of $S$ depends only on $\varepsilon$ and $t$, not on $x$ or $y$.

\textbf{Controlling term (2).}
With $S$ now fixed, \cref{lem:V-lipschitz-continuous-x-finite-T} gives that $V_S(t,\cdot)$ is Lipschitz continuous in the state with some finite constant $L_{Vx}$. Therefore:
\[
|V_S(t,x)-V_S(t,y)|
\leq
L_{Vx} \|x-y\|.
\]

Choose
\[
\delta = \frac{\varepsilon}{3L_{Vx}}.
\]
Then whenever $\|x-y\| < \delta$:
\[
|V_S(t,x)-V_S(t,y)|
<
\frac{\varepsilon}{3}.
\]

\textbf{Combining.}
For this choice of $\delta$, whenever $\|x-y\| < \delta$:
\[
|V(t,x)-V(t,y)|
<
\frac{\varepsilon}{3}
+
\frac{\varepsilon}{3}
+
\frac{\varepsilon}{3}
=
\varepsilon.
\]

Since $\varepsilon$ was arbitrary, $V(t,\cdot)$ is continuous at $x$. Since $x$ was arbitrary, $V(t,\cdot)$ is continuous on $\mathbb{R}^n$.
\end{proof}

\begin{boxedlemma}[Lipschitz Continuity of $V(t,x)$ in $t$ and $W(\tau, x)$ in $\tau$ for $T$ Finite]
\label{lem:V-lipschitz-continuous-t-infinite-T}
For every fixed $x\in\mathbb{R}^n$ and finite $T<\infty$: (i) the time-to-go value function $W(\tau,x)$ (\cref{def:W}) is Lipschitz continuous in $\tau$ on $[0,T]$, with some constant $K_{x,T}>0$; (ii) consequently, the value function $V(t,x)$ (\cref{def:V}) is Lipschitz continuous in $t$ on $[0,T]$ with the same constant.
\end{boxedlemma}

\begin{proof}
We first prove the result for the time-to-go value function \cref{def:W}.
Fix \(x\in\mathbb R^n\) and let \(T<\infty\). Since
\(u\mapsto f(x,u)\) is continuous and \(\mathcal{U}\) is compact, the quantity
\[
C_x:=\sup_{u\in \mathcal{U}}\|f(x,u)\|
\]
is finite.

Let \(z(r)=\phi_x^u(r)\). By global Lipschitz continuity of \(f\) in
the state variable, uniformly in \(u\), we have
\[
\begin{aligned}
\|f(z(r),u(r))\|
&\le
\|f(z(r),u(r))-f(x,u(r))\|
+
\|f(x,u(r))\| \\
&\le
L_f\|z(r)-x\|+C_x .
\end{aligned}
\]
Therefore,
\[
\|z(r)-x\|
\le
\int_0^r L_f\|z(q)-x\|\,dq+C_x r .
\]
By Gr\"onwall's inequality,
\[
\|z(r)-x\|
\le
C_x r e^{L_f r},
\qquad r\in[0,T].
\]
Consequently, for all \(r\in[0,T]\),
\[
\|f(z(r),u(r))\|
\le
C_x+L_f C_x T e^{L_fT}.
\]
Define
\[
C_{x,T}:=C_x+L_f C_x T e^{L_fT}.
\]
Then, for any \(r,q\in[0,T]\), without loss of generality \(q\le r\),
\[
\begin{aligned}
\|\phi_x^u(r)-\phi_x^u(q)\|
&=
\left\|
\int_q^r f(\phi_x^u(\ell),u(\ell))\,d\ell
\right\| \\
&\le
\int_q^r \|f(\phi_x^u(\ell),u(\ell))\|\,d\ell \\
&\le
C_{x,T}|r-q|.
\end{aligned}
\]
Thus,
\[
\|\phi_x^u(r)-\phi_x^u(q)\|
\le
C_{x,T}|r-q|
\]
for all \(r,q\in[0,T]\), uniformly over all admissible controls.

Now define, for each admissible control \(u(\cdot)\),
\[
F_u(r):=e^{-\lambda r}g(\phi_x^u(r)).
\]
Using boundedness of \(g\), Lipschitz continuity of \(g\), and the
trajectory estimate above, we obtain
\[
\begin{aligned}
|F_u(r)-F_u(q)|
&=
\left|
e^{-\lambda r}g(\phi_x^u(r))
-
e^{-\lambda q}g(\phi_x^u(q))
\right| \\
&\le
\left|
e^{-\lambda r}g(\phi_x^u(r))
-
e^{-\lambda q}g(\phi_x^u(r))
\right| \\
&\quad
+
\left|
e^{-\lambda q}g(\phi_x^u(r))
-
e^{-\lambda q}g(\phi_x^u(q))
\right| \\
&\le
M|e^{-\lambda r}-e^{-\lambda q}|
+
L_g\|\phi_x^u(r)-\phi_x^u(q)\| \\
&\le
\left(M\lambda+L_g C_{x,T}\right)|r-q|.
\end{aligned}
\]
Let
\[
K_{x,T}:=M\lambda+L_g C_{x,T}.
\]
Then
\[
|F_u(r)-F_u(q)|
\le
K_{x,T}|r-q|
\]
for all \(r,q\in[0,T]\), uniformly over admissible controls.

We now show that \(W(\cdot,x)\) is Lipschitz continuous. Let
\[
0\le \tau_1\le \tau_2\le T,
\qquad
\eta:=\tau_2-\tau_1.
\]
Since \([0,\tau_1]\subset[0,\tau_2]\), increasing the horizon can only
decrease the infimum. Hence
\[
W(\tau_2,x)\le W(\tau_1,x).
\]
We prove the reverse inequality up to an error of order \(\eta\).

Fix an arbitrary control \(u\in\mathcal U[0,\tau_2]\). Define
\[
A_u:=\inf_{r\in[0,\tau_1]}F_u(r),
\qquad
B_u:=\inf_{r\in[\tau_1,\tau_2]}F_u(r).
\]
For every \(r\in[\tau_1,\tau_2]\), the uniform Lipschitz estimate for
\(F_u\) gives
\[
F_u(r)
\ge
F_u(\tau_1)-K_{x,T}\eta.
\]
Since
\[
F_u(\tau_1)\ge A_u,
\]
we obtain
\[
B_u\ge A_u-K_{x,T}\eta.
\]
Therefore,
\[
\inf_{r\in[0,\tau_2]}F_u(r)
=
\min\{A_u,B_u\}
\ge
A_u-K_{x,T}\eta.
\]
The restriction of \(u\) to \([0,\tau_1]\) is an admissible control for
the horizon \(\tau_1\). Hence
\[
A_u\ge W(\tau_1,x).
\]
Thus,
\[
\inf_{r\in[0,\tau_2]}F_u(r)
\ge
W(\tau_1,x)-K_{x,T}\eta.
\]
Taking the infimum over all \(u\in\mathcal U[0,\tau_2]\) yields
\[
W(\tau_2,x)
\ge
W(\tau_1,x)-K_{x,T}\eta.
\]
Combining this with \(W(\tau_2,x)\le W(\tau_1,x)\), we get
\[
0
\le
W(\tau_1,x)-W(\tau_2,x)
\le
K_{x,T}(\tau_2-\tau_1).
\]
Therefore,
\[
|W(\tau_2,x)-W(\tau_1,x)|
\le
K_{x,T}|\tau_2-\tau_1|.
\]
Hence \(W(\cdot,x)\) is Lipschitz continuous on \([0,T]\).

For finite terminal time \(T<\infty\), the absolute-time value function
satisfies
\[
V(t,x)=W(T-t,x).
\]
Therefore, for any \(t_1,t_2\in[0,T]\),
\[
\begin{aligned}
|V(t_2,x)-V(t_1,x)|
&=
|W(T-t_2,x)-W(T-t_1,x)| \\
&\le
K_{x,T}|(T-t_2)-(T-t_1)| \\
&=
K_{x,T}|t_2-t_1|.
\end{aligned}
\]
Thus \(V(\cdot,x)\) is Lipschitz continuous in \(t\), and therefore
continuous.
\end{proof}

\begin{boxedlemma}[Pointwise Continuity of $V^{\infty}(t,x)$ in $t$]
        \label{lem:V-continuous-t-infinite}
        For every fixed $x \in \mathbb{R}^n$, the value function $V^{\infty}(t,x)$ (\cref{def:V-infinity}) is pointwise continuous in $t$.
\end{boxedlemma}

\begin{proof}
    \(V^{\infty}(t,x)\) has been proven to be invariant with respect to $t$ (\cref{lem:V-infinity-t-invariant}). Therefore, \(V^{\infty}(\cdot,x)\) is continuous in $t$.
\end{proof}

\subsection{Dynamic Programming Principle (DPP)}

The DPP associated with the avoid case has already been established in \cite{choi2021robust}. Following similar proof mechanics, we establish the DPP associated with the reach problem.

\begin{boxedproposition}[DPP Associated with $W(\tau, x)$]
    \label{prop:dpp-W}
    Consider the Bellman step $\sigma > 0$, then the DPP associated with $W(\tau,x)$  (\cref{def:W}) is as follows.
    \begin{equation}
        \begin{aligned}
            W(\tau, x)&= \inf_{u \in \mathcal{U}[0, \sigma]} \min \{\underbrace{\inf _{r \in[0, \sigma]} g\left(\phi_x^u(r)\right) e^{-\lambda r}}_{\text{stopping branch}}, \\
            &\underbrace{e^{-\lambda \sigma} W\left(\tau-\sigma, \phi_{x}^{u}(\sigma)\right)}_{\text{continuation branch}}\}
        \end{aligned}
        \label{eq:W_DPP}
    \end{equation}
\end{boxedproposition}

\begin{proof}
Based on \cref{eq:W_eqn}, 
\begin{equation}
    W(\tau, x)=\inf _{u(\cdot) \in \mathcal{U}[0, \tau]} \inf _{r \in[0, \tau]} g(\phi_{x}^{u}(r)) e^{-\lambda r}.
\end{equation}

Based on $[0, \tau] = [0, \sigma]\cup [\sigma, \tau]$, and by admissibility and concatenation of controls, we may optimize independently:
\begin{equation}
    \begin{aligned}
    W(\tau, x) &= \inf_{u_1 \in \mathcal{U}[0, \sigma]} \inf_{u_2 \in \mathcal{U}[\sigma, \tau]}\min \Big \{\inf_{r \in [0, \sigma]} g(\phi_{x}^{u_1}(r)) e^{-\lambda r}, \\
    &\inf_{r \in [\sigma, \tau]} g(\phi_{\phi_x^{u_1}(\sigma)}^{u_2}(r)) e^{-\lambda r} \Big\}
    \end{aligned}
\end{equation}

For the second term, let $q:=r-\sigma$. Thus, we get
\begin{equation}
\begin{aligned}
    & \inf_{q \in [0, \tau-\sigma]} g(\phi_{\phi_x^{u_1}(\sigma)}^{u_2}(r)) e^{-\lambda (q + \sigma)}\\
    &=e^{-\lambda \sigma} \inf_{q \in [0, \tau-\sigma]}  g(\phi_{\phi_x^{u_1}(\sigma)}^{u_2}(q+\sigma)) e^{-\lambda q} 
\end{aligned}
\end{equation}

Since the term on the left-hand side only depends on $u_1$, we can move the $\inf_{u_{2} \in \mathcal{U}[\sigma, \tau]}$ inside to the term on the right-hand side,
\begin{equation}
    \begin{aligned}
    W(\tau, x) &= \inf_{u_1 \in \mathcal{U}[0, \sigma]} \min \Big \{ \inf_{r \in [0,\sigma]} g(\phi_{x}^{u_1}(r)) e^{-\lambda r}, \\
    &  e^{-\lambda \sigma} \inf_{u_2 \in \mathcal{U}[\sigma, \tau]} \inf_{q \in [0, \tau-\sigma]}  g(\phi_{\phi_x^{u_1}(\sigma)}^{u_2}(q+\sigma)) e^{-\lambda q} \Big\}
    \end{aligned}
\end{equation}

Through semigroup flow property, the term on the right-hand side becomes
\begin{equation}
\begin{aligned} 
    &e^{-\lambda \sigma} \inf_{u_2 \in \mathcal{U}[\sigma, \tau]} \inf_{q \in [0, \tau-\sigma]}  g(\phi_{\phi_{x}^{u_1}(\sigma)}^{u_2}(q)) e^{-\lambda q}\\
    &=e^{-\lambda \sigma} W(\tau - \sigma,\phi_{x}^{u_1}(\sigma))
\end{aligned}
\end{equation}

Putting it all together (and renaming $u_1$ to $u$ so that it is more generic) yields \cref{eq:W_DPP}.
\end{proof}

\begin{boxedproposition}[DPP Associated with $W^{\infty}(x)$]
    \label{prop:dpp-W-infinity}
        \begin{equation}
        \begin{aligned}
            W^{\infty}(x)&= \inf_{u \in \mathcal{U}[0, \sigma]} \min \{\underbrace{\inf _{r \in[0, \sigma]} g\left(\phi_x^u(r)\right) e^{-\lambda r}}_{\text{stopping branch}}, \\
            &\underbrace{e^{-\lambda \sigma} W^{\infty}\left(\phi_{x}^{u}(\sigma)\right)}_{\text{continuation branch}}\}
        \end{aligned}
        \label{eq:W-infinity_DPP}
    \end{equation}
\end{boxedproposition}

\begin{proof}
The result follows directly from \cref{eq:W_DPP} by taking $T \to \infty$.  Since $W(\tau, x) \to W^{\infty}(x)$ as $\tau \to \infty$ (\cref{def:W-inf}), and the Bellman step $\sigma > 0$ is fixed and finite, the interval $[0, \sigma]$  is unaffected by the limit. Substituting $W^{\infty}$ for $W(\tau - \sigma, \varphi^u_x(\sigma))$ in \cref{eq:W_DPP} yields \cref{eq:W-infinity_DPP}.
\end{proof}

\subsection{Reachability Encoding}

We briefly clarify how the proposed value function encodes reachability. In HJ reachability, it is standard to represent reachable sets via the sign structure of a value function defined in terms of a terminal cost $g$ \cite{lygeros2004reachability,bardi1997optimal}. In particular, for reach problems, $g(x) < 0$ for $x$ in the target set and $g(x) \geq 0$ otherwise (\cref{ass:g-sign-calibration}), so that the negative sublevel set of the value function characterizes the backward reachable tube.

\begin{boxedproposition}[Reachability encoding]
Let $V$ be the value function from \cref{def:V}, and suppose $g(x) < 0$ if and only if $x \in \mathcal G$. Then, the strict backward reachable tube associated with $\mathcal G$ is given by
\begin{equation*}
    \mathcal R = \{ x \in \mathcal X : V(x) < 0 \}.
\end{equation*}
\end{boxedproposition}
\begin{proof}
    The result follows directly from the definition of $V$ (\cref{def:V}) and the sign-preserving property of the discount factor.
    Since
    \begin{equation}
    e^{-\lambda (s-t)} > 0,
    \end{equation}
    it follows that
    \begin{equation}
    \operatorname{sign}\!\left(g(\xi(s))\, e^{-\lambda (s-t)}\right)
    =
    \operatorname{sign}\!\left(g(\xi(s))\right).
    \end{equation}
    
    Therefore,
    \begin{equation}
    \begin{aligned}
        &\inf_{s \ge t} g(\xi(s))\, e^{-\lambda (s-t)} < 0 \\
        \iff &\exists\, s \ge t \text{ such that } g(\xi(s)) < 0 \\
        \iff &x \in \mathcal{R}_{\mathrm{BRT}}.
    \end{aligned}
    \end{equation}
\end{proof}

\section{HJ Characterization}
\label{sec:hj-characterization}

We show that the discounted reach-based value function $W(\tau,x)$, and its infinite-horizon counterpart $W^\infty(x)$, are each the unique bounded continuous viscosity solution of an associated Hamilton--Jacobi variational inequality (HJVI). The finite-horizon case is treated in \cref{thm:W-unique-soln-HJVI}, and the infinite-horizon case in \cref{thm:W-unique-soln-HJVI-infinity}.

\subsection{HJ Viscosity Characterization of $W(\tau,x)$}

\begin{boxedtheorem}[HJ Viscosity Characterization of $W(\tau,x)$]
\label{thm:W-unique-soln-HJVI}
$W(\tau,x)$ (\cref{eq:W_eqn}) is a unique bounded continuous viscosity solution of:
    \begin{equation}
        \label{eq:HJVI}
        \begin{aligned}
            \min \Big \{&W_{\text{HJ}}(\tau, x) - g(x), \frac{\partial W_{\text{HJ}}}{\partial \tau}(\tau,x) \\
            &- \inf_{u \in \mathcal{U}} \frac{\partial W_{\text{HJ}}(\tau, x)}{\partial x} f(x, u) + \lambda W_{\text{HJ}}(\tau,x)\Big\} = 0, \\[10pt]
        \end{aligned}
    \end{equation}
\end{boxedtheorem}

\begin{proof}
    The proof follows standard arguments in viscosity theory for HJ variational inequalities (see e.g. Bardi \& Capuzzo-Dolcetta \cite{bardi1997optimal}), adapted to the discounted reach-stopping structure. The result (adapted for the avoid case) is corroborated by Theorem 3 in \cite{choi2021robust}.
\end{proof}

\subsection{HJ Viscosity Characterization of $W^{\infty}(x)$}

\begin{boxedtheorem}[HJ Viscosity Characterization of $W^{\infty}(x)$]
    \label{thm:W-unique-soln-HJVI-infinity}
    $W^{\infty}(x)$ is the unique bounded continuous stationary viscosity solution of
    \begin{equation}
        \label{eq:HJVI-infinity}
        \begin{aligned}
            \min \Big \{&W^{\infty}(x) - g(x), \\
            &\lambda W^{\infty}(x) - \inf_{u \in \mathcal{U}} \frac{\partial W^{\infty}(x)}{\partial x} f(x, u)\Big\} = 0
        \end{aligned}
    \end{equation}
\end{boxedtheorem}

\begin{proof}
The comparison principle and viscosity characterization for eq.~\eqref{eq:HJVI-infinity}
follow the same standard arguments as \cref{thm:W-unique-soln-HJVI}, restricted to
the stationary (time-independent) setting. In particular, since $W^\infty(x)$ is
independent of $\tau$ (\cref{def:W-inf}), it satisfies the same viscosity
sub- and supersolution conditions as $W(\tau,x)$ with the $\tau$-derivative term
vanishing, so that \cref{eq:HJVI} reduces exactly to
\cref{eq:HJVI-infinity}. Comparison for the resulting stationary equation is
classical (see e.g.\ \cite[Ch.~II--III]{bardi1997optimal})
\end{proof}

\section{Reachability-Preserving Bellman Operator}
\label{sec:bellman-exact}

Up to this point, we have relied on the discounted reachability value function introduced by Choi et al.~\cite{choi2021robust} and its associated HJ characterization. The missing ingredient is a Bellman operator whose fixed point characterizes exactly the same value function. This section derives such a reachability-preserving Bellman operator, for both the finite-horizon time-to-go value function $W(\tau,x)$ and its infinite-horizon counterpart $W^{\infty}(x)$, with each operator following directly from the corresponding DPP (\cref{prop:dpp-W} and \cref{prop:dpp-W-infinity}). We also establish the conditions for Banach's fixed-point theorem to apply, which in turn means that each Bellman operator admits a unique fixed point (\cref{thm:unique-fixed-point-T} and \cref{thm:unique-fixed-point-T-infinity}).

\subsection{Finite-Horizon Bellman Operator \texorpdfstring{$\mathcal{T}_{\sigma, \lambda}$}{}}
\label{subsec:finite-horizon-bellman-op}

The Bellman operator associated with $W(\tau, x)$ can be inferred from the established DPP for $W(\tau,x)$ (\cref{prop:dpp-W}). It is presented in the following.

\begin{boxeddefinition}[Bellman Operator Associated with $W(\tau,x)$]
\label{def:T-finite}
For a bounded function $\Psi$

\begin{equation}
    \Psi: [0, T] \times \mathbb{R}^{n} \to \mathbb{R},
\end{equation}
and for a chosen discount rate $\lambda > 0$ and a chosen Bellman step $\sigma > 0$, we define the Bellman operator, denoted by $\mathcal{T}_{\sigma, \lambda}$, by:
    \begin{equation}
        \begin{aligned}
            (\mathcal T_{\sigma,\lambda}\Psi)(\tau,x)
            := \\
            \inf_{u(\cdot)\in\mathcal U[0,h(\tau)]}
            \min\Bigg\{
            &
            \inf_{r\in[0,h(\tau)]}
            e^{-\lambda r}g(\phi_x^u(r)),
            \\
            &
            e^{-\lambda h(\tau)}
            \Psi\bigl(\tau-h(\tau),\phi_x^u(h(\tau))\bigr)
            \Bigg\}, \\
            &\text{where} \ h(\tau) = \min\{\tau,\sigma\}
        \end{aligned}
        \label{eq:T}
    \end{equation}
We also define the \emph{discount factor} as $\gamma:= e^{-\lambda \sigma}$.
\end{boxeddefinition}

\begin{boxedremark}[Role of $h(\tau)$]
    The function \(h(\tau)\) (see \cref{eq:T}) ensures that the Bellman update never rolls out beyond the remaining time horizon. In particular, if \(\tau<\sigma\), then only \(\tau\) units of time remain, and the operator must use the interval \([0,\tau]\), not \([0,\sigma]\).
\end{boxedremark}

\subsection{Finite-Horizon Bellman Operator Unique Fixed Point}
\label{subsec:finite-horizon-bellman-unique-fixed-point}

We now proceed to prove that the finite-horizon Bellman operator (\cref{def:T-finite}) admits a unique fixed-point.  Banach's fixed-point theorem will be used to establish the existence and uniqueness of the fixed point of $\mathcal{T}_{\sigma,\lambda}$. This requires showing that the operator maps a Banach space of value functions into itself (\cref{lem:finite-horizon-invariance}) and is contractive under the supremum norm. These properties are established in the following.

The finite-horizon time-to-go value function $W(\tau,x)$ satisfies the boundary condition
\[
W(0,x)=g(x).
\]
Therefore, we define a function space that has this boundary condition directly built into it:

\begin{boxeddefinition}[Space of Value Functions $\mathcal{B}$]
    \label{def:B}
    Let
    \begin{equation}
        \label{eq:set-V}
        \begin{aligned}
            \mathcal V &:= \Big\{\Psi: [0,T]\times\mathbb R^n \to \mathbb{R},\\
            &\|\Psi\|_{\infty} < \infty, \Psi(0, x) = g(x) \qquad \forall x \in \mathbb{R}^n \Big \}
        \end{aligned}
    \end{equation}
    denote the set of bounded functions that satisfy the boundary condition $\Psi(0,x) = g(x)$, where
    \[
    \|\Psi\|_\infty
    :=
    \sup_{(\tau,x)\in[0,T]\times\mathbb R^n}
    |\Psi(\tau,x)|.
    \]
    
    Let $\mathcal{B}$ denote $\mathcal{V}$ equipped with the metric: \[
        d(\Psi_1,\Psi_2):=\|\Psi_1-\Psi_2\|_\infty.
    \]
    Thus,
        \[
        \mathcal B :=
        (\mathcal V,d).
        \]
\end{boxeddefinition}

\begin{boxedlemma}[\(\mathcal B\) Is Complete]
\label{lem:B-complete}
The space \(\mathcal B:= (\mathcal V,d)\) is complete.
\end{boxedlemma}

\begin{proof}
Let \((\Psi_k)_{k\in\mathbb N}\) be a Cauchy sequence in
\((\mathcal V,d)\). Since each \(\Psi_k\) is bounded and the space of
bounded real-valued functions on \([0,T]\times\mathbb R^n\) is complete under
the supremum norm, there exists a bounded function
\[
\Psi:[0,T]\times\mathbb R^n\to\mathbb R
\]
such that
\[
\|\Psi_k-\Psi\|_\infty\to0.
\]
It remains to verify the boundary condition. For every \(x\in\mathbb R^n\),
\[
\Psi(0,x)
=
\lim_{k\to\infty}\Psi_k(0,x)
=
g(x),
\]
because \(\Psi_k\in\mathcal V\) for every \(k\). Hence
\[
\Psi\in\mathcal V.
\]
Therefore \(\mathcal V\) is a closed subset of the complete metric space of
bounded functions. Hence \(\mathcal V\) is complete.
\end{proof}

\begin{boxedlemma}[Invariance of $\mathcal V$ Under $\mathcal T_{\sigma,\lambda}$]
\label{lem:finite-horizon-invariance}
Let
\[
\Psi\in\mathcal V.
\]
Then
\[
\mathcal T_{\sigma,\lambda}\Psi\in\mathcal V.
\]
Equivalently,
\[
\mathcal T_{\sigma,\lambda}(\mathcal V)
\subseteq
\mathcal V.
\]
\end{boxedlemma}

\begin{proof}
We prove boundedness and the preservation of the boundary condition.\\

\textbf{Boundedness.}

Since $\Psi \in \mathcal V$,
\[
\|\Psi\|_\infty < \infty.
\]

From \cref{ass:g-bounded}, the uniform boundedness means that pointwise boundedness also holds:
\begin{equation}
    |g(\phi_{x}^{u}(r))| \leq M \qquad \forall r \in [0, \sigma], u \in \mathcal{U}][0, \sigma]
\end{equation}

Since $r>0$ and $\lambda>0$, $e^{-\lambda r} \in (0, 1)$. Thus,
\begin{equation}
    |g(\phi_{x}^{u}(r)) e^{-\lambda r}| \leq M \qquad \forall r \in [0, \sigma], u \in \mathcal{U}][0, \sigma]
\end{equation}

Therefore
\[
-M
\le
\inf_{r\in[0,\sigma]}
g(\phi_x^u(r))e^{-\lambda r}
\le
M, \qquad \forall u \in \mathcal{U}][0, \sigma]
\]
and hence
\[
\left|
\inf_{r\in[0,\sigma]}
g(\phi_x^u(r))e^{-\lambda r}
\right|
\le M \qquad \forall u \in \mathcal{U}][0, \sigma].
\]

Moreover,
\[
|e^{-\lambda\sigma}
\Psi(\tau-h(\tau),\phi_x^u(\sigma))|
\le
\|\Psi\|_\infty .
\]

Thus both branches are bounded by
\[
\max\{M,\|\Psi\|_\infty\}.
\]

It follows that
\[
|(\mathcal T_{\sigma,\lambda}\Psi)(\tau,x)|
\le
\max\{M,\|\Psi\|_\infty\}
\]

for all $(\tau,x)$. Taking the supremum over $(\tau,x)$,
\[
\|\mathcal T_{\sigma,\lambda}\Psi\|_\infty
\le
\max\{M,\|\Psi\|_\infty\}
<\infty.
\]

Hence $\mathcal T_{\sigma,\lambda}\Psi$ is bounded.

\textbf{Boundary condition} It remains to verify the boundary condition. At \(\tau=0\),
\[
h(0)=0.
\]
Therefore,
\[
\inf_{r\in[0,h(0)]}
e^{-\lambda r}g(\phi_x^u(r))
=
g(x),
\]
and
\[
e^{-\lambda h(0)}
\Psi\bigl(0-h(0),\phi_x^u(h(0))\bigr)
=
\Psi(0,x)
=
g(x),
\]
because \(\Psi\in\mathcal V\). Hence
\[
(\mathcal T_{\sigma,\lambda}\Psi)(0,x)=g(x).
\]
Therefore
\[
\mathcal T_{\sigma,\lambda}\Psi\in\mathcal V.
\]
\end{proof}

\begin{boxedlemma}[Sup-Norm Contraction Property of $\mathcal{T}_{\sigma, \lambda}$]
    \label{lem:contraction}
    For all $\Psi_1,\Psi_2\in\mathcal V$.  
    \begin{equation}
        \big\|\mathcal{T}_{\sigma, \lambda} \Psi_1-\mathcal{T}_{\sigma, \lambda} \Psi_2\big\|_{\infty} \leq e^{-\lambda \sigma} \left\|\Psi_1-\Psi_2\right\|_{\infty} .
    \end{equation}
    \label{lem:contraction-T}
\end{boxedlemma}

\begin{proof}
    Using \cref{def:T-finite} and applying the non-expansiveness of the infimum,
    \begin{equation}
    \begin{aligned}
        &\left|\mathcal{T}_{\sigma, \lambda} \Psi_1(\tau, x)-\mathcal{T}_{\sigma, \lambda} \Psi_2(\tau, x)\right| \leq \\
        &\sup_{u \in \mathcal{U}[0,\sigma]} \Big| \min \Big\{\inf _{r \in[0, \sigma]} g\left(\phi_x^u(r)\right) e^{-\lambda r}, e^{-\lambda \sigma} \Psi_1\left(\tau-\sigma, \phi_x^u(\sigma)\right)\Big\} \\
        & - \min \Big\{\inf _{r \in[0, \sigma]} g\left(\phi_x^u(r)\right) e^{-\lambda r}, e^{-\lambda \sigma} \Psi_2\left(\tau-\sigma, \phi_x^u(\sigma)\right)\Big\} \Big|
    \end{aligned}
    \end{equation}
    
    Moreover, the minimum operator is 1-Lipschitz. This means that
    \begin{equation}
        |\min (a, c)-\min (b, c)| \leq|a-b| .
    \end{equation}

    Therefore,
    \begin{equation}
    \begin{aligned}
        &\left|\mathcal{T}_{\sigma, \lambda} \Psi_1(\tau, x)-\mathcal{T}_{\sigma, \lambda} \Psi_2(\tau, x)\right| \leq \\
        &\sup_{u \in \mathcal{U}[0,\sigma]} \Big|e^{-\lambda \sigma} \Psi_1\left(\tau-\sigma, \phi_x^u(\sigma)\right) - e^{-\lambda \sigma} \Psi_2\left(\tau-\sigma, \phi_x^u(\sigma)\right)  \Big|
    \end{aligned}
    \end{equation}

    We may pull $e^{-\lambda \sigma}$ outside since it is independent of $u$ and is a positive coefficient:
    \begin{equation}
        \begin{aligned}
        & \left|\mathcal{T}_{\sigma, \lambda} \Psi_1(\tau, x)-\mathcal{T}_{\sigma, \lambda} \Psi_2(\tau, x)\right| \\
        & \leq e^{-\lambda \sigma} \sup_{u(\cdot) \in \mathcal{U}[0, \sigma]} \Big|\Psi_1\left(\tau-\sigma, \phi_x^u(\sigma)\right)-\Psi_2\left(\tau-\sigma, \phi_x^u(\sigma)\right)\Big| .
        \end{aligned}
    \end{equation}
    By the definition of the sup-norm
    \begin{equation}
        \left|\mathcal{T}_{\sigma, \lambda} \Psi_1(\tau, x)-\mathcal{T}_{\sigma, \lambda} \Psi_2(\tau, x)\right| \leq e^{-\lambda \sigma} \left\|\Psi_1-\Psi_2\right\|_{\infty}
    \end{equation}
    Taking the supremum over $(\tau, x)$ on both sides preserves the inequality, since the supremum is order-preserving. This concludes the proof.
\end{proof}

\begin{boxedtheorem}[Existence and Uniqueness of the Fixed Point of $\mathcal{T}_{\sigma, \lambda}$]
\label{thm:unique-fixed-point-T}
The finite-horizon Bellman operator $\mathcal T_{\sigma,\lambda}$ admits a unique fixed point in $\mathcal{V}$, denoted by $W_{\text{Bell}}$:
\[
W_{\text{Bell}} \in\mathcal V.
\]
such that
\[
W_{\text{Bell}} = \mathcal{T}_{\sigma,\lambda} W_{\text{Bell}}
\]
Moreover, for any initial $\Psi_0 \in \mathcal{V}$, the iteration
\[
\Psi_{k+1} = \mathcal{T}_{\sigma,\lambda} \Psi{k}
\]
converges uniformly to $W_{\text{Bell}}$, and
\[
\| \Psi_k - W_{\text{Bell}}\|_{\infty} \leq e^{-k \lambda \sigma} \| \Psi_0 - W_{\text{Bell}}\|_{\infty}
\]
\end{boxedtheorem}

\begin{proof}
By \cref{lem:B-complete},
\(\mathcal B\) is a complete space. By \cref{lem:finite-horizon-invariance},
\(\mathcal T_{\sigma,\lambda}\) maps \(\mathcal V\) into itself. By \cref{lem:contraction}, \(\mathcal T_{\sigma,\lambda}\) is a contraction with contraction factor \(e^{-\lambda\sigma}<1\). Thus, the conditions for Banach's fixed-point theorem are satisfied. The theorem, consequently, implies that
\(\mathcal T_{\sigma,\lambda}\) admits a unique fixed point
\(W_{\text{Bell}}\in\mathcal V\). The theorem also implies that, for every
\(\Psi_0\in\mathcal V\), the Picard iteration
\[
\Psi_{k+1}=\mathcal T_{\sigma,\lambda}\Psi_k
\]
converges uniformly to \(W\), with
\[
\|\Psi_k-W_{\text{Bell}}\|_\infty
\le
e^{-k\lambda\sigma}
\|\Psi_0-W_{\text{Bell}}\|_\infty.
\]

\end{proof}

\begin{boxedtheorem}[$W = W_{\mathrm{Bell}}$]
\label{thm:W-equals-Wbell}
The discounted reachability value function $W$ from \cref{def:W} 
is the unique fixed point of $\mathcal{T}_{\sigma,\lambda}$ in 
$\mathcal{B}$, i.e. $W = W_{\mathrm{Bell}}$.
\end{boxedtheorem}
\begin{proof}
By the DPP (\cref{prop:dpp-W}) and \cref{def:T-finite}, 
$W$ satisfies $W = \mathcal{T}_{\sigma,\lambda}W$, so $W$ is a fixed point. 
Uniqueness follows from \cref{thm:unique-fixed-point-T}. Hence, $W = W_{\mathrm{Bell}}$.
\end{proof}

\subsection{Infinite-Horizon Bellman Operator \texorpdfstring{$\mathcal{T}^{\infty}_{\sigma, \lambda}$}{}}
\label{subsec:infinite-horizon-bellman-op}

We now state the corresponding infinite-horizon result. In the infinite-horizon
case, since the dynamics, the cost function \(g\), and the control set \(\mathcal{U}\) are invariant in time, the value function becomes stationary.
That is,
\[
V^\infty(t,x)=W^\infty(x),
\]
where
\[
W^{\infty}(x)
:=
\inf_{u(\cdot)\in\mathcal U[0,\infty)}
\inf_{r\in[0,\infty)}
e^{-\lambda r}g(\phi_x^u(r)).
\]
The stationary problem has no time-to-go variable and therefore no boundary
condition \(W(0,x)=g(x)\).

\begin{boxeddefinition}[Stationary Infinite-Horizon Bellman Operator $\mathcal T^\infty_{\sigma,\lambda}$]
\label{def:T-infinity}
For \(\Psi\in \mathcal{B}^{\infty}\), define
\[
\begin{aligned}
(\mathcal T^\infty_{\sigma,\lambda}\Psi)(x)
:=
\inf_{u(\cdot)\in\mathcal U[0,\sigma]}
\min\Bigg\{
&
\inf_{r\in[0,\sigma]}
e^{-\lambda r}g(\phi_x^u(r)),
\\
&
e^{-\lambda\sigma}\Psi(\phi_x^u(\sigma))
\Bigg\}.
\end{aligned}
\]
\end{boxeddefinition}

\subsection{Infinite-Horizon Bellman Operator Unique Fixed Point}

\begin{boxeddefinition}[Space of Stationary Value Functions $\mathcal{B}^{\infty}$]
\label{def:B-infinity}
Let
\begin{equation}
\label{eq:set-V-infinity}
    \mathcal{V}^{\infty}
    :=
    \left\{
    \Psi:\mathbb R^n\to\mathbb R
    \ \middle|\
    \|\Psi\|_\infty<\infty
    \right\},
\end{equation}
where
\[
\|\Psi\|_\infty:=\sup_{x\in\mathbb R^n}|\Psi(x)|.
\]

Let $\mathcal{B}^{\infty}$ be the function space where $\mathcal{V}^{\infty}$ is equipped with the metric: \[
    d(\Psi_1,\Psi_2):=\|\Psi_1-\Psi_2\|_\infty.
\]
i.e.
    \[
    \mathcal B^{\infty} :=
    (\mathcal V^{\infty},d).
    \]
\end{boxeddefinition}

\begin{boxedlemma}[$\mathcal{B}^{\infty}$ Is Complete]
    \label{lem:B-inf-complete}
        \[
    \mathcal B^{\infty} :=
    (\mathcal V^{\infty},d)
    \]
    is a complete space.
\end{boxedlemma}
\begin{proof}
    The proof follows the same structure as that of \cref{lem:B-complete}, without needing to look into the boundary condition as it is not relevant for the infinite horizon case.
\end{proof}

\begin{boxedlemma}[Invariance of $\mathcal{V}^{\infty}$ Under $\mathcal{T}_{\sigma,\lambda}^{\infty}$]
    \label{lem:infinite-horizon-invariance}
\end{boxedlemma}

\begin{proof}
    In this case, only the boundedness needs to be proven since the boundary condition no longer is relevant for the infinite-horizon case.
    Let \(\Psi\in \mathcal{B}^{\infty}\). Since \(g\) is bounded by \(M\),
    \[
    \left|
    e^{-\lambda r}g(\phi_x^u(r))
    \right|
    \le
    M
    \]
    for every \(r\in[0,\sigma]\), every \(x\), and every admissible control.
    Also,
    \[
    |e^{-\lambda\sigma}\Psi(\phi_x^u(\sigma))|
    \le
    \|\Psi\|_\infty.
    \]
    Therefore,
    \[
    |(\mathcal T^\infty_{\sigma,\lambda}\Psi)(x)|
    \le
    \max\{M,\|\Psi\|_\infty\}.
    \]
    Taking the supremum over \(x\), we get
    \[
    \|\mathcal T^\infty_{\sigma,\lambda}\Psi\|_\infty
    \le
    \max\{M,\|\Psi\|_\infty\}<\infty.
    \]
    Thus
    \[
    \mathcal T^\infty_{\sigma,\lambda}\Psi\in \mathcal{B}^{\infty}.
    \]
\end{proof}

\begin{boxedlemma}[Sup-Norm Contraction Property of $\mathcal T^\infty_{\sigma,\lambda}$]
\label{lem:contraction-T-infinity}
For all $\Psi_1, \Psi_2 \in \mathcal{V}^{\infty}$
\[
\|\mathcal T^\infty_{\sigma,\lambda}\Psi_1
-
\mathcal T^\infty_{\sigma,\lambda}\Psi_2\|_\infty
\le
e^{-\lambda\sigma}
\|\Psi_1-\Psi_2\|_\infty
\]
\end{boxedlemma}

\begin{proof}
Let \(\Psi_1,\Psi_2\in \mathcal{B}^{\infty}\). Using the non-expansiveness of
the infimum and the 1-Lipschitz property of the minimum operator,
\[
\begin{aligned}
&
|(\mathcal T^\infty_{\sigma,\lambda}\Psi_1)(x)
-
(\mathcal T^\infty_{\sigma,\lambda}\Psi_2)(x)|
\\
&\le
\sup_{u(\cdot)\in\mathcal U[0,\sigma]}
\left|
e^{-\lambda\sigma}\Psi_1(\phi_x^u(\sigma))
-
e^{-\lambda\sigma}\Psi_2(\phi_x^u(\sigma))
\right|
\\
&\le
e^{-\lambda\sigma}
\|\Psi_1-\Psi_2\|_\infty.
\end{aligned}
\]
Taking the supremum over \(x\) proves the contraction property.
\end{proof}

\begin{boxedtheorem}[Existence and Uniqueness of the Fixed Point of $\mathcal{T}^{\infty}_{\sigma,\lambda}$]
\label{thm:unique-fixed-point-T-infinity}
Let \(\lambda>0\) and \(\sigma>0\). The stationary Bellman operator
\(\mathcal T^\infty_{\sigma,\lambda}\) admits a unique fixed point
\[
W_{\text{Bell}}^{\infty}\in \mathcal{B}^{\infty}.
\]
Moreover, for any initial \(\Psi_0\in \mathcal{B}^{\infty}\), the iteration
\[
\Psi_{k+1}
=
\mathcal T^\infty_{\sigma,\lambda}\Psi_k
\]
converges uniformly to \(W^{\infty}\), with
\[
\|\Psi_k-W_{\text{Bell}}^{\infty}\|_\infty
\le
e^{-k\lambda\sigma}
\|\Psi_0-W_{\text{Bell}}^{\infty}\|_\infty.
\]
\end{boxedtheorem}

\begin{proof}
The space \(\mathcal{B}^{\infty}\) is complete (\cref{lem:B-inf-complete}).
Moreover, by \cref{lem:infinite-horizon-invariance},
\[
\mathcal T^\infty_{\sigma,\lambda}:\mathcal{B}^{\infty}\to \mathcal{B}^{\infty}
\]
and by \cref{lem:contraction-T-infinity},
\[
\|\mathcal T^\infty_{\sigma,\lambda}\Psi_1
-
\mathcal T^\infty_{\sigma,\lambda}\Psi_2\|_\infty
\le
e^{-\lambda\sigma}
\|\Psi_1-\Psi_2\|_\infty.
\]
Since \(\lambda>0\) and \(\sigma>0\),
\[
e^{-\lambda\sigma}<1.
\]
Therefore, Banach's fixed-point theorem gives a unique fixed point
\(W^{\infty}\in \mathcal{B}^{\infty}\), and the Picard iteration converges
uniformly to it with the stated geometric rate.
\end{proof}

\begin{boxedtheorem}[$W^\infty = W_{\mathrm{Bell}}^{\infty}$]
\label{thm:W-inf-equals-Wbell-inf}
The infinite-horizon discounted reachability value function $W^{\infty}$ from \cref{def:W-inf} is the unique fixed point of $\mathcal{T}^{\infty}_{\sigma,\lambda}$ in $\mathcal{B}^\infty$, i.e. $W^{\infty} = W_{\mathrm{Bell}}^{\infty}$.
\end{boxedtheorem}
\begin{proof}
By the DPP (\cref{prop:dpp-W-infinity}) and \cref{def:T-infinity}, $W^{\infty}$ satisfies
\[
W^{\infty}(x) = (\mathcal{T}^{\infty}_{\sigma,\lambda} W^{\infty})(x) 
\qquad \forall x \in \mathbb{R}^n,
\]
so $W^{\infty}$ is a fixed point of $\mathcal{T}^{\infty}_{\sigma,\lambda}$. Uniqueness follows from \cref{thm:unique-fixed-point-T-infinity}. Hence, $W^{\infty} = W_{\mathrm{Bell}}^{\infty}$.
\end{proof}



\begin{boxedremark}[Finite horizon versus infinite horizon]
The finite-horizon operator acts on functions of \((\tau,x)\) and requires the
boundary condition
\[
\Psi(0,x)=g(x).
\]
This boundary condition is essential for obtaining a uniform contraction when
\(\tau<\sigma\).

By contrast, the infinite-horizon operator is stationary and acts only on
functions of \(x\). It has no terminal or boundary condition, and the
contraction follows directly from the discounted continuation term
\(e^{-\lambda\sigma}\Psi(\phi_x^u(\sigma))\).
\end{boxedremark}

\begin{boxedremark}[Convergence speed of value iteration and numerical consistency tradeoff]
    As seen from \cref{thm:unique-fixed-point-T} and \cref{thm:unique-fixed-point-T-infinity}, $\sigma$ influences convergence speed. For example, if $\sigma$ is small, then the value iteration convergence becomes slow. On the other hand, numerical consistency often favours small $\sigma$. Therefore, a tradeoff between contraction speed and numerical consistency exists, and it is influenced by the choice of $\sigma$.
\end{boxedremark}

\section{Equivalence of Characterizations}
\label{sec:equivalence}

Having established both the HJ characterization (\cref{sec:hj-characterization}) and the Bellman fixed-point characterization (\cref{sec:bellman-exact}), we now show that these two characterizations are equivalent and identify the same value function. We establish this equivalence for both the finite-horizon and infinite-horizon settings.

\subsection{Finite-Horizon Equivalence of Characterizations}

\begin{boxedtheorem}[Finite-Horizon HJ--Bellman Equivalence]
\label{thm:equivalence-characterization}
The HJ and Bellman characterizations of the discounted reachability value
function coincide in the finite horizon setting.
\end{boxedtheorem}

\begin{proof}

The discounted reachability value function $W$ from \cref{def:W} satisfies:
\begin{enumerate}
    \item $W$ is the unique bounded continuous viscosity solution of
    \cref{eq:HJVI} (\cref{thm:W-unique-soln-HJVI}).
    \item $W = W_{\mathrm{Bell}}$: $W$ is the unique fixed point of 
    $\mathcal{T}_{\sigma,\lambda}$ in $\mathcal{B}$ 
    (\cref{thm:unique-fixed-point-T}).
\end{enumerate}
Thus, $W = W_{\mathrm{Bell}}$. Hence, the HJ and Bellman characterizations are equivalent.

\end{proof}

\subsection{Infinite-Horizon Equivalence of Characterizations}
\begin{boxedtheorem}[Infinite-Horizon HJ--Bellman Equivalence]
\label{thm:equivalence-characterization-infinity}
The HJ and Bellman characterizations of the discounted reachability value function coincide in the infinite horizon setting.
\end{boxedtheorem}
\begin{proof}
The discounted reachability value function $W^{\infty}$ from \cref{def:W-inf}
satisfies:
\begin{enumerate}
    \item $W^{\infty}$ is the unique bounded continuous viscosity solution of
    \cref{eq:HJVI-infinity} (\cref{thm:W-unique-soln-HJVI-infinity}).
    \item $W^{\infty} = W^{\infty}_{\mathrm{Bell}}$: $W^{\infty}$ is the unique
    fixed point of $\mathcal{T}^{\infty}_{\sigma,\lambda}$ in $\mathcal{B}^{\infty}$
    (\cref{thm:unique-fixed-point-T-infinity}).
\end{enumerate}
Combining (1) and (2), $W^{\infty}$ is simultaneously the unique viscosity
solution of \cref{eq:HJVI-infinity} and the unique fixed point of
$\mathcal{T}^{\infty}_{\sigma,\lambda}$. Hence the HJ and Bellman
characterizations are equivalent in the infinite-horizon setting.
\end{proof}

\section{Computational Realization of the Bellman Operator}
\label{sec:bellman-numerical}

The Bellman operator introduced in \cref{sec:bellman-exact} characterizes the discounted reachability value function as the unique fixed point of a contraction mapping. In this section, we translate this fixed-point characterization into a computational procedure. We first construct a discrete approximation of the operator for each of the finite and infinite horizon cases in \cref{subsec:discrete-bellman-operator,subsec:discrete-bellman-operator-infinity}. Their convergences, in the limit of the discretization parameters going to zero, to their respective true operators is shown in \cref{subsec:convergence-numerical-finite-horizon,subsec:convergence-numerical-infinite-horizon}. 

\subsection{Finite-Horizon Discrete Bellman Operator $\widehat{\mathcal{T}}_{\sigma,\lambda}$}
\label{subsec:discrete-bellman-operator}
We now construct a discrete (in time) approximation of the Bellman operator suitable for computation.

We restrict the time-to-go variable to the discrete time grid:
\[
\tau_k = k\sigma,
\]
and define
\begin{equation}
    W_k(x) := W(\tau_k,x) \quad \text{where} \quad  W_{0}(x) = g(x) .
\end{equation}
Evaluating the Bellman operator (\cref{def:T-finite}) at $\tau = \tau_{k} = k\sigma$ yields    
\begin{equation}
\label{eq:Wk}
\begin{aligned}
W_{k}(x)
=
&\inf _{u(\cdot) \in \mathcal{U}[0, \sigma]} \min\!\Big\{\inf _{r \in[0, \sigma]} g\left(\phi_x^u(r)\right) e^{-\lambda r},\\
&e^{-\lambda \sigma}
W_{k-1}\!\left(\phi_x^u(\sigma)\right)
\Big\}.
\end{aligned}
\end{equation}

To implement the recursion numerically, the exact flow $\phi_x^u(\cdot)$ is replaced by a consistent numerical approximation. Let
\[
\widehat F_{\Delta t} : \mathbb R^n \times \mathcal U \to \mathbb R^n
\]
denote a one-step numerical integrator for the dynamics $\dot x = f(x,u)$ over a small time step $\Delta t$.

For example, explicit Euler yields
\[
\widehat F_{\Delta t}(x,u) = x + \Delta t \, f(x,u),
\]
while higher-order Runge--Kutta methods provide more accurate approximations. In general,
\[
\widehat F_{\Delta t}(x,u) = \phi_x^u(\Delta t) + \mathcal O(\Delta t^p)
\]
for some order $p \ge 1$.

We approximate the flow over a Bellman interval $[0,\sigma]$ by composing $H = \sigma / \Delta t$ steps. We denote by $\widehat{F}_\sigma(x, u)$ the $H$-fold composition of $\widehat{F}_{\Delta t}$:
\[
\widehat{F}_\sigma(x, u) := \underbrace{(\widehat{F}_{\Delta t} \circ \cdots \circ \widehat{F}_{\Delta t})}_{H \text { times }}(x, u).
\]

\begin{boxedremark}[Separation of time scales]
    The Bellman discretization step $\sigma$ plays a distinct role from the numerical integration step $\Delta t$ used to approximate trajectories. The Bellman operator requires evaluating the minimum of the discounted reach cost over the interval $r \in [0,\sigma]$. To approximate this quantity, we introduce a finer time discretization $\Delta t \ll \sigma$ to resolve trajectories within each Bellman update. This separation is essential for consistency with the continuous-time formulation.
\end{boxedremark}

\begin{boxeddefinition}[Fully Discrete Bellman Operator $\widehat{\mathcal T}_{\sigma,\lambda}$]
\label{def:T-hat}
Let $\Psi \in \mathcal{V}$ (see \cref{eq:set-V}).

Let $\widehat F_{\Delta t}$ be a numerical flow map, and let $\mathcal U_d$ be a finite discretization of the control set. Let $r_j = j \Delta t$ for $j = 0, \dots, H$, where $H = \sigma / \Delta t$.

In the discrete approximation, the control input $u$ is assumed piecewise-constant over the Bellman interval (zero-order hold parameterization).

The fully discrete Bellman operator $\widehat{\mathcal T}_{\sigma,\lambda}$ acting on a value function $\Psi$ is defined by
\begin{equation}
\label{eq:fully-discrete-operator-correct}
\begin{aligned}
(\widehat{\mathcal T}_{\sigma,\lambda}\Psi)(\tau_k, x)
&= \min_{u \in \mathcal U_d} \min \Big\{\underbrace{\min_{j=0,\dots,H} g\!\big(x_j^u\big) e^{-\lambda r_j}}_{\text{stopping branch}}, \\
&\underbrace{e^{-\lambda\sigma} \Psi\!\big(\tau_{k} - h(\tau_{k}),x_H^u\big)}_{\text{continuation branch}}\Big\},\\
\end{aligned}
\end{equation}
where $h(\tau_{k}) = \min\{\tau_{k},\sigma\}$ and the trajectory $\{x_j^u\}$ is defined recursively by
\[
x_0^u = x, \quad x_{j+1}^u = \widehat F_{\Delta t}(x_j^u, u).
\]
\end{boxeddefinition}

\subsection{Infinite-Horizon Discrete Bellman Operator $\widehat{\mathcal{T}}^{\infty}_{\sigma,\lambda}$}
\label{subsec:discrete-bellman-operator-infinity}

In the infinite horizon case, the value function becomes independent of $\tau$. Thus, only the control space is discretized.

\begin{boxeddefinition}[Fully Discrete Bellman Operator $\widehat{\mathcal T}_{\sigma,\lambda}^{\infty}$ for Infinite Horizon]
\label{def:T-hat-infinity}
Let $\Psi \in \mathcal{B}^\infty$ (see \cref{def:B-infinity}). 
Let $\widehat{F}_{\Delta t}$ be a numerical flow map, and let $\mathcal{U}_d$ be a finite discretization 
of the control set $\mathcal{U}$. Let $r_j = j \Delta t$ for $j = 0, \dots, H$, where $H = \sigma/\Delta t$.
As in the finite-horizon case, the control input $u$ is assumed piecewise-constant over the Bellman
interval (zero-order hold parameterization).

The fully discrete Bellman operator $\widehat{\mathcal T}^{\infty}_{\sigma,\lambda}$ acting on a value function $\Psi$ is defined by
\begin{equation}
\label{eq:T-hat-infinity}
\begin{aligned}
(\widehat{\mathcal T}^{\infty}_{\sigma,\lambda}\Psi)(x)
&= \min_{u \in \mathcal U_d} \min \Big\{\underbrace{\min_{j=0,\dots,H} g\!\big(x_j^u\big) e^{-\lambda r_j}}_{\text{stopping branch}}, \\
&\underbrace{e^{-\lambda\sigma} \Psi\!\big(x_H^u\big)}_{\text{continuation branch}}\Big\},\\
\end{aligned}
\end{equation}
where the trajectory $\{x^u_j\}$ is defined recursively by
\[
x^u_0 = x, \qquad x^u_{j+1} = \widehat{F}_{\Delta t}(x^u_j, u).
\]
\end{boxeddefinition}

\subsection{Convergence of the Numerical Scheme for Finite Horizon}
\label{subsec:convergence-numerical-finite-horizon}

In this subsection, we prove the convergence of the numerical scheme. Namely, we prove the convergence of $\widehat W$ to $W$, where $\widehat W$ is the fixed point of $\widehat{\mathcal{T}}_{\sigma,\lambda}$ (\cref{def:T-hat}), i.e.
\begin{equation}
    \label{eq:W-hat}
    \widehat W = \widehat{\mathcal{T}}_{\sigma,\lambda} \widehat W .
\end{equation}
We also prove the convergence of $\widehat W^{\infty}$ to $W^{\infty}$, where $\widehat W^{\infty}$ is the fixed point of $\widehat{\mathcal T}^{\infty}_{\sigma,\lambda}$ (\cref{def:T-hat-infinity}).

\begin{boxedlemma}[Consistency of $\widehat{\mathcal{T}}_{\sigma,\lambda}$]
\label{lem:consistency-T-hat}
Assume that the numerical flow map
$\widehat F_{\Delta t}$ is a consistent approximation of the
exact flow $\phi_x^u$, i.e.,
\[
\widehat F_{\Delta t}(x,u)
=
\phi_x^u(\Delta t)
+
O(\Delta t^p)
\]
for some order $p\ge 1$.

Assume furthermore that the control discretization
$\mathcal{U}_d$ converges to $\mathcal{U}$ as the discretization is refined.

Then, for $\Psi \in \mathcal{V}$ (see \cref{eq:set-V}), $\widehat{\mathcal{T}}_{\sigma,\lambda}$ converges pointwise to the Bellman operator $\mathcal{T}_{\sigma,\lambda}$ (\cref{def:T-finite}):

\[
(\widehat{\mathcal{T}}_{\sigma,\lambda}\Psi)(\tau_k,x)
\rightarrow
(\mathcal{T}_{\sigma,\lambda}\Psi)(\tau_k,x)
\]

as

\[
\Delta t \rightarrow 0,
\qquad
\mathcal{U}_d \rightarrow U,
\]

for every fixed Bellman step $\sigma > 0$.
\end{boxedlemma}

\begin{proof}
Consistency of the numerical integrator implies

\[
x_j^u
=
\phi_x^u(r_j)
+
O(\Delta t^p).
\]

Since $g$ is continuous, as $\Delta t \to 0$

\[
g(x_j^u)e^{-\lambda r_j}
\rightarrow
g(\phi_x^u(r_j))e^{-\lambda r_j}.
\]

As $\Delta t\to0$, the discrete minimum

\[
\min_{j=0,\ldots,H}
g(x_j^u)e^{-\lambda r_j}
\]

converges to

\[
\inf_{r\in[0,\sigma]}
g(\phi_x^u(r))e^{-\lambda r}.
\]

Likewise,

\[
e^{-\lambda\sigma}
\Psi(\tau_{k-1},x_H^u)
\rightarrow
e^{-\lambda\sigma}
\Psi(\tau-\sigma,\phi_x^u(\sigma)).
\]

Taking minima over controls and combining both terms yields

\[
\widehat{\mathcal{T}}_{\sigma,\lambda}\Psi
\rightarrow
\mathcal{T}_{\sigma,\lambda}\Psi.
\]

Moreover, since the Lipschitz constants $L_{fx}$ (\cref{ass:f-lipschitz-x}) and
$L_g$ (\cref{ass:g-lipschitz-x}) are independent of $x$ and $u$, the constant
implicit in the $O(\Delta t^p)$ bound above depends only on $\sigma$,
$L_{fx}$, $L_g$, and the integrator order $p$ --- not on $x$, $u$, or
$\tau_k$. Hence
\[
x_j^u \to \phi_x^u(r_j)
\]
uniformly over $x \in \mathbb{R}^n$ and $u \in \mathcal{U}_d$, and consequently
\[
(\widehat{T}_{\sigma,\lambda}\Psi)(\tau_k, \cdot) \to
(T_{\sigma,\lambda}\Psi)(\tau_k, \cdot)
\]
uniformly on $\mathbb{R}^n$, i.e.\ in the sup norm, for every fixed
$\sigma > 0$.
\end{proof}

\begin{boxedlemma}[Sup-Norm Contraction Property of $\widehat{\mathcal{T}}_{\sigma, \lambda}$]
    Let $\Psi_1,\Psi_2\in\mathcal V$.  
    \begin{equation}
        \big\|\widehat{\mathcal{T}}_{\sigma, \lambda} \Psi_1-\widehat{\mathcal{T}}_{\sigma, \lambda} \Psi_2\big\|_{\infty} \leq e^{-\lambda \sigma} \left\|\Psi_1-\Psi_2\right\|_{\infty} .
    \end{equation}
    \label{lem:contraction-T-hat}
\end{boxedlemma}

\begin{proof}
    The proof can be established using the same mechanics used in the proof of \cref{lem:contraction-T}.
\end{proof}

\begin{boxedtheorem}[Convergence of $\widehat W$ to $W$]   
    \label{thm:What-conv-W}
    Let
    \begin{equation*}
        W = \mathcal{T}_{\sigma, \lambda} W
    \end{equation*}
    and let
    \begin{equation*}
        \widehat W = \widehat{\mathcal{T}}_{\sigma, \lambda} \widehat W
    \end{equation*}
    Then
    \begin{equation*}
        \widehat W \to W \qquad \text{as } \Delta t \to 0
    \end{equation*}
\end{boxedtheorem}

\begin{proof}
    \begin{equation}
        \|\widehat{W} - W \|_{\infty} = \|\widehat{\mathcal{T}}_{\sigma,\lambda} \widehat W - \mathcal{T}_{\sigma,\lambda} W\|_{\infty}
    \end{equation}

    Now we add and subtract $\widehat{\mathcal{T}}_{\sigma,\lambda} W$

    \begin{equation}
        \|\widehat{\mathcal{T}}_{\sigma,\lambda} \widehat W - \mathcal{T}_{\sigma,\lambda} W + \widehat{\mathcal{T}}_{\sigma,\lambda} W - \widehat{\mathcal{T}}_{\sigma,\lambda} W \|_{\infty}
    \end{equation}

    And apply the triangle inequality
    \begin{equation}
    \begin{aligned}
        &\|\widehat{\mathcal{T}}_{\sigma,\lambda} \widehat W - \mathcal{T}_{\sigma,\lambda} W + \widehat{\mathcal{T}}_{\sigma,\lambda} W - \widehat{\mathcal{T}}_{\sigma,\lambda} W \|_{\infty}\\
        & \leq \|\widehat{\mathcal{T}}_{\sigma,\lambda} (\widehat W - W)\|_{\infty} + \|\widehat{\mathcal{T}}_{\sigma,\lambda} W  - \mathcal{T}_{\sigma,\lambda} W \|_{\infty}
    \end{aligned}
    \end{equation}

    By the contraction property of $\widehat{\mathcal{T}}_{\sigma,\lambda}$
    \begin{equation}
        \|\widehat{\mathcal{T}}_{\sigma,\lambda} (\widehat W - W)\|_{\infty}  \leq e^{-\lambda \sigma} \|\widehat{W} - W \|_{\infty}
    \end{equation}

    Thus,
    \begin{equation}
        \|\widehat{W} - W \|_{\infty} \leq e^{-\lambda \sigma} \|\widehat{W} - W \|_{\infty} + \|\widehat{\mathcal{T}}_{\sigma,\lambda} W  - \mathcal{T}_{\sigma,\lambda} W \|_{\infty}
    \end{equation}

    Rearranging leads to:
    \begin{equation}
        (1 - e^{-\lambda \sigma}) \|\widehat{W} - W \|_{\infty} \leq \|\widehat{\mathcal{T}}_{\sigma,\lambda} W  - \mathcal{T}_{\sigma,\lambda} W \|_{\infty}
    \end{equation}
    Since \cref{lem:consistency-T-hat} gives that $\|\widehat{\mathcal{T}}_{\sigma,\lambda} W  - \mathcal{T}_{\sigma,\lambda} W \|_{\infty} \to 0$, this concludes the proof.
\end{proof}

\subsection{Convergence of the Numerical Scheme for Infinite Horizon}
\label{subsec:convergence-numerical-infinite-horizon}

We now establish the analogous convergence result for the infinite-horizon discrete Bellman operator $\widehat{\mathcal{T}}^{\infty}_{\sigma,\lambda}$, following similar proof mechanics to those utilized in \cref{subsec:convergence-numerical-finite-horizon}.

\begin{boxedlemma}[Consistency of $\widehat{\mathcal{T}}^{\infty}_{\sigma,\lambda}$]
\label{lem:consistency-T-hat-infinity}
Assume that the numerical flow map
$\widehat F_{\Delta t}$ is a consistent approximation of the
exact flow $\phi_x^u$, i.e.,
\[
\widehat F_{\Delta t}(x,u)
=
\phi_x^u(\Delta t)
+
O(\Delta t^p)
\]
for some order $p\ge 1$.

Assume furthermore that the control discretization
$\mathcal{U}_d$ converges to $\mathcal{U}$ as the discretization is refined.

Then, for $\Psi \in \mathcal{V}^{\infty}$ (see \cref{eq:set-V-infinity}), $\widehat{\mathcal{T}}^{\infty}_{\sigma,\lambda}$ converges pointwise to the Bellman operator $\mathcal{T}^{\infty}_{\sigma,\lambda}$ (\cref{def:T-infinity}):

\[
(\widehat{\mathcal{T}}^{\infty}_{\sigma,\lambda}\Psi)(x)
\rightarrow
(\mathcal{T}^{\infty}_{\sigma,\lambda}\Psi)(x)
\]

as

\[
\Delta t \rightarrow 0,
\qquad
\mathcal{U}_d \rightarrow U,
\]

for every fixed Bellman step $\sigma > 0$.
\end{boxedlemma}

\begin{proof}
The argument follows the same mechanics as the proof of \cref{lem:consistency-T-hat}, with the 
time-remaining bookkeeping ($\tau_{k}$, $h(\tau_{k})$) omitted, since the 
stationary operator has no time-to-go argument.

Consistency of the numerical integrator implies
\[
x^u_j = \phi^u_x(r_j) + O(\Delta t^p).
\]
Since $g$ is continuous, as $\Delta t \to 0$
\[
g(x^u_j)e^{-\lambda r_j} \to g(\phi^u_x(r_j))e^{-\lambda r_j}.
\]
As $\Delta t \to 0$, the discrete minimum
\[
\min_{j=0,\dots,H} g(x^u_j)e^{-\lambda r_j}
\]
converges to
\[
\inf_{r\in[0,\sigma]} g(\phi^u_x(r))e^{-\lambda r}.
\]
Likewise, since $x^u_H \to \phi^u_x(\sigma)$ and $\Psi$ is continuous,
\[
e^{-\lambda\sigma}\Psi(x^u_H) \to e^{-\lambda\sigma}\Psi(\phi^u_x(\sigma)).
\]
Taking minima over controls and combining both terms yields
$\widehat{T}^\infty_{\sigma,\lambda}\Psi \to T^\infty_{\sigma,\lambda}\Psi$.
\end{proof}

\begin{boxedlemma}[Sup-Norm Contraction Property of $\widehat{\mathcal{T}}^{\infty}_{\sigma, \lambda}$]
    \label{lem:contraction-T-hat-infinity}
    Let $\Psi_1,\Psi_2\in\mathcal V^{\infty}$ (\cref{eq:set-V-infinity}).  
    \begin{equation}
        \big\|\widehat{\mathcal{T}}^{\infty}_{\sigma, \lambda} \Psi_1-\widehat{\mathcal{T}}^{\infty}_{\sigma, \lambda} \Psi_2\big\|_{\infty} \leq e^{-\lambda \sigma} \left\|\Psi_1-\Psi_2\right\|_{\infty} .
    \end{equation}
\end{boxedlemma}

\begin{proof}
    The proof can be established using the same mechanics used in the proof of \cref{lem:contraction-T-infinity}.
\end{proof}

\begin{boxedtheorem}[Convergence of $\widehat W^{\infty}$ to $W^{\infty}$]   
    \label{thm:What-conv-W-infinity}
    Let
    \begin{equation*}
        W^{\infty} = \mathcal{T}^{\infty}_{\sigma, \lambda} W^{\infty}
    \end{equation*}
    and let
    \begin{equation*}
        \widehat W^{\infty} = \widehat{\mathcal{T}}^{\infty}_{\sigma, \lambda} \widehat W^{\infty}
    \end{equation*}
    Then
    \begin{equation*}
        \widehat W^{\infty} \to W^{\infty} \qquad \text{as } \Delta t \to 0
    \end{equation*}
\end{boxedtheorem}

\begin{proof}
    The proof relies on similar mechanics as in the case of \cref{thm:What-conv-W} and again leveraging the contraction property, now of $\widehat{T}^{\infty}_{\sigma,\lambda}$.
\end{proof}

\section{Reinforcement Learning as a Sample-Based Approximation of the Bellman Operator}
\label{sec:rl-sampled-based}

The discrete Bellman operators $\widehat{T}_{\sigma,\lambda}$ and
$\widehat{T}^{\infty}_{\sigma,\lambda}$ introduced in the previous
subsections provide computable approximations of the exact operators,
whose fixed points converge to $W$ and $W^{\infty}$ respectively as
$\Delta t \to 0$ (\cref{thm:What-conv-W}). However, even these discrete
formulations require exact rollouts of the system dynamics and
exhaustive optimization over $\mathcal{U}_d$, which may be intractable
in high-dimensional settings or when the dynamics are unknown. Instead,
the operators can be approximated using sampled rollouts, leading to a
sample-based approximation of value iteration in which Bellman updates
are constructed from data rather than exact evaluations.

We make this connection concrete by considering fitted value iteration
(FVI), a value-based, sample-based method that approximates the fixed
point of the reachability-preserving Bellman operator via bootstrapped
regression. We note that FVI, as presented here, is more precisely
described as an instance of \emph{approximate dynamic programming}
(ADP)~\cite{bertsekas1996neuro}: the Bellman targets are computed
via exact rollouts of a known dynamics model rather than through
interaction with an unknown environment, which is the setting more
commonly associated with RL. The connection to RL
becomes direct once the rollout and optimization steps are themselves
replaced by sampled trajectories and learned policies, an extension we
discuss briefly at the end of this section.

Since the finite-horizon and infinite-horizon Bellman operators act on
functions of different domains ($[0,T]\times\mathbb{R}^n$ versus
$\mathbb{R}^n$, respectively), we introduce two corresponding parametric
approximators. Let
\begin{equation}
    \Psi_\theta : [0,T] \times \mathbb{R}^n \to \mathbb{R}
\end{equation}
denote the finite-horizon parametric approximator, and let
\begin{equation}
    \Psi_\theta^{\infty} : \mathbb{R}^n \to \mathbb{R}
\end{equation}
denote its infinite-horizon (stationary) counterpart, where $n$ denotes
the dimension of the state. In both cases, the goal is to minimize the
Bellman residual using sampled states. Given a dataset $\{x_i\}_{i=1}^N$
(where the samples are obtained from a sampling distribution $\rho$),
the parameters $\theta$ are updated by minimizing, in the finite-horizon
case,
\begin{equation}
    \mathcal{L}(\theta) = \mathbb{E}_{(\tau,x)\sim \rho}
    \left[\left(\Psi_\theta(\tau,x) -
    \underbrace{\widehat{T}_{\sigma,\lambda}(\Psi_\theta)(\tau,x)}_{y(\tau,x)}
    \right)^2\right],
    \label{eq:fvi-loss}
\end{equation}
and, in the infinite-horizon case,
\begin{equation}
    \mathcal{L}^\infty(\theta) = \mathbb{E}_{x\sim \rho}
    \left[\left(\Psi_\theta^\infty(x) -
    \underbrace{\widehat{T}^\infty_{\sigma,\lambda}(\Psi_\theta^\infty)(x)}_{y^\infty(x)}
    \right)^2\right].
\end{equation}

For each sampled state, the target is determined by a sampled
evaluation of the corresponding discrete Bellman operator. In the
finite-horizon case (based on \cref{eq:fully-discrete-operator-correct}),
\begin{equation}
\begin{aligned}
    y(\tau,x) &= \min_{u \in \mathcal U_d} \min \Big\{\min_{j=0,\dots,H} g(x_j^u)\,\\
    &e^{-\lambda r_j},\;e^{-\lambda\sigma}\, \Psi_{\theta}(\tau - \sigma,\, x_H^u)\Big\},
\end{aligned}
\label{eq:fvi-target}
\end{equation}
and, in the infinite-horizon case,
\begin{equation}
y^\infty(x) =
\min_{u \in \mathcal U_d}
\min \left\{
\min_{j=0,\dots,H} g(x_j^u)\,e^{-\lambda r_j},
\;
e^{-\lambda\sigma}\, \Psi_{\theta}^\infty(x_H^u)
\right\}.
\end{equation}

Algorithms~\ref{alg:fvi-finite} and~\ref{alg:fvi-infinite} give a realization of the numerical scheme for the finite-horizon and infinite-horizon settings, respectively.

\begin{algorithm}
\caption{Finite-Horizon FVI with $\widehat{T}_{\sigma,\lambda}$}
\label{alg:fvi-finite}
\begin{algorithmic}[1]
\Require value function $\Psi_\theta(\tau, \cdot)$ structure, reach cost $g$, control set 
$\mathcal{U}_d$, dynamics model, optimizer $\mathcal{O}$, sampling distribution $\rho$ over 
$[0,T] \times \mathbb{R}^n$, batch size $B$, Bellman step $\sigma$, timestep $\Delta t$, 
discount rate $\lambda$, pretraining conververgence threshold $\epsilon_{\text{pre-training}}$, convergence threshold $\epsilon$ 

\State Initialize parameters $\theta$

\Statex \textit{Pretraining on reach cost (enforces boundary condition $\Psi_\theta(0, x) = g(x)$)}
\For{$k = 1$ to $N_\text{pre}$}
    \State Sample $\{x_i\}_{i=1}^B \sim \rho(\cdot \mid \tau = 0)$
    \State $y_i \leftarrow g(x_i)$
    \State $\mathcal{L}_\text{pre}(\theta) \leftarrow \frac{1}{B} \sum_{i=1}^B \left(\Psi_\theta(0, x_i) - y_i\right)^2$
    \State $\theta \leftarrow \mathcal{O}(\theta, \mathcal{L}_\text{pre}(\theta))$
\EndFor

\Statex \textit{Fitted value iteration}
\While{$\frac{1}{B}\sum_{i=1}^B \left(\Psi_\theta(\tau_i, x_i) - \widehat{T}_{\sigma,\lambda}(\Psi_\theta)(\tau_i, x_i)\right)^2 > \epsilon$}
    \State Sample $\{(\tau_i, x_i)\}_{i=1}^B \sim \rho$
    \For{each $(\tau_i, x_i)$}
        \State Set $h_i \leftarrow \min\{\tau_i, \sigma\}$, $H_i \leftarrow h_i / \Delta t$
        \For{each $u \in \mathcal{U}_d$}
            \State Roll out trajectory $\{x^u_j\}_{j=0}^{H_i}$ using dynamics
            \State $J_\text{stop}(u) \leftarrow \min_{j=0,\ldots,H_i} e^{-\lambda j \Delta t} g(x^u_j)$
            \State $J_\text{cont}(u) \leftarrow e^{-\lambda h_i} \Psi_\theta(\tau_i - h_i,\, x^u_{H_i})$
            \State $J(u) \leftarrow \min(J_\text{stop}(u),\, J_\text{cont}(u))$
        \EndFor
        \State $y_i \leftarrow \min_{u \in \mathcal{U}_d} J(u)$
    \EndFor
    \State $\mathcal{L}(\theta) \leftarrow \frac{1}{B} \sum_{i=1}^B \left(\Psi_\theta(\tau_i, x_i) - y_i\right)^2$
    \State $\theta \leftarrow \mathcal{O}(\theta, \mathcal{L}(\theta))$
\EndWhile
\State \Return $\Psi_\theta$
\end{algorithmic}
\end{algorithm}

\begin{algorithm}
\caption{Infinite-Horizon FVI with $\widehat{T}^\infty_{\sigma,\lambda}$}
\label{alg:fvi-infinite}
\begin{algorithmic}[1]
\Require value function $\Psi^{\infty}_\theta(x)$ structure, reach cost $g$, control set $\mathcal{U}_d$, 
dynamics model, optimizer $\mathcal{O}$, sampling distribution $\rho$ over $\mathbb{R}^n$, 
batch size $B$, Bellman step $\sigma$, timestep $\Delta t$, discount rate $\lambda$, 
convergence threshold $\epsilon$

\State Initialize parameters $\theta$

\Statex \textit{Fitted value iteration}
\While{$\frac{1}{B}\sum_{i=1}^B \left(\Psi^{\infty}_\theta(x_i) - \widehat{T}^\infty_{\sigma,\lambda}(\Psi^{\infty}_\theta)(x_i)\right)^2 > \epsilon$}
    \State Sample $\{x_i\}_{i=1}^B \sim \rho$
    \For{each $x_i$}
        \For{each $u \in \mathcal{U}_d$}
            \State Roll out trajectory $\{x^u_j\}_{j=0}^{H}$ using dynamics, where $H = \sigma/\Delta t$
            \State $J_\text{stop}(u) \leftarrow \min_{j=0,\ldots,H} e^{-\lambda j \Delta t} g(x^u_j)$
            \State $J_\text{cont}(u) \leftarrow e^{-\lambda\sigma} \Psi^{\infty}_\theta(x^u_H)$
            \State $J(u) \leftarrow \min(J_\text{stop}(u),\, J_\text{cont}(u))$
        \EndFor
        \State $y_i \leftarrow \min_{u \in \mathcal{U}_d} J(u)$
    \EndFor
    \State $\mathcal{L}^{\infty}(\theta) \leftarrow \frac{1}{B} \sum_{i=1}^B \left(\Psi^{\infty}_\theta(x_i) - y_i\right)^2$
    \State $\theta \leftarrow \mathcal{O}(\theta, \mathcal{L}^{\infty}(\theta))$
\EndWhile
\State \Return $\Psi_\theta^{\infty}$
\end{algorithmic}
\end{algorithm}

\begin{boxedremark}[Relationship between \cref{alg:fvi-finite} and \cref{alg:fvi-infinite}]
\label{rem:fvi-alg-relationship}
\Cref{alg:fvi-infinite} specializes \cref{alg:fvi-finite} to the
stationary setting in three ways: (i) pretraining is omitted since
$\mathcal{V}^\infty$ imposes no boundary condition (see
\cref{def:B-infinity}), and Banach's theorem guarantees convergence
from any bounded initialization; (ii) rollouts always run the full
$H = \sigma/\Delta t$ steps since there is no remaining time
constraint, whereas in \cref{alg:fvi-finite} rollouts are truncated to
$H_i = h_i/\Delta t$ steps when $\tau_i < \sigma$; and (iii) the
network $\Psi_\theta^{\infty}$ takes only $x$ as input, reflecting the
stationarity of $W^\infty(x)$, in contrast to $\Psi_\theta(\tau, x)$ in
the finite-horizon case.
\end{boxedremark}

Having established both the finite- and infinite-horizon FVI schemes as sample-based realizations of their respective discrete Bellman operators, we close with a caveat regarding what has, and has not, been established about their convergence in \cref{rem:fvi-convergence}.

\begin{boxedremark}[From Discrete to Learned Operators: An Open Convergence Question]
\label{rem:fvi-convergence}
\Cref{lem:consistency-T-hat,lem:consistency-T-hat-infinity} establish
that the discrete operators
$\widehat{T}_{\sigma,\lambda}$ and $\widehat{T}^{\infty}_{\sigma,\lambda}$
recover the exact operators $T_{\sigma,\lambda}$ and
$T^{\infty}_{\sigma,\lambda}$ as discretization parameters vanish. This
concerns the discretization of trajectories and controls, not the
introduction of function approximation.

The FVI scheme of \cref{sec:rl-sampled-based} adds a further approximation, composing the discrete operator with a least-squares projection onto a parametric class $\{\Psi_\theta\}$ (respectively $\{\Psi_\theta^\infty\}$; \cref{eq:fvi-loss,eq:fvi-target}). Whether the resulting iterates converge to the fixed point $\widehat{W}$ (respectively $\widehat{W}^{\infty}$) is a distinct question: the contraction property of
\cref{lem:contraction-T,lem:contraction-T-infinity,lem:contraction-T-hat,,lem:contraction-T-hat-infinity} need not be inherited by the composite map ``regress-then-apply-operator''. Bounding or ruling out the resulting error falls within the scope of the approximate dynamic programming literature on FVI~\cite{bertsekas1996neuro,munos2005error,munos2008finite}, which propagates function-approximation and finite-sample error through the Bellman recursion for standard, additive-reward operators. Extending such guarantees to the non-additive, min-structured operator introduced here is left for future work (see \cref{sec:conclusion}).
\end{boxedremark}

\section{Experiments \& Results}
\label{sec:exp-res}

The objective of the experiments is to empirically support the key theoretical claims of this work, namely: (i) the preservation of reachability semantics under discounting, (ii) the equivalence between the HJ and Bellman characterizations of the discounted reachability value function, and (iii) the interpretation of RL algorithms as sample-based approximation schemes for the fixed-point of the proposed Bellman operator.

\subsection{Experimental Setup}

\paragraph{Double Integrator Reach Problem}

We consider the reach problem applied to the double integrator, a canonical time-invariant control system for which HJ reachability structure is well understood. The dynamics and control set are given by \cref{eq:di_dyn}.
\begin{equation}
\dot x_{1} = x_{2},\qquad \dot x_{2} = u,\qquad u\in\mathcal U:=[-u_{\text{max}}, u_{\text{max}}],
\label{eq:di_dyn}
\end{equation}
where $u_{\text{max}}$ is the maximum control that can be provided.

In order to satisfy \cref{ass:g-bounded,ass:g-sign-calibration}, the reach cost $g(x)$ is chosen as a scaled and squashed signed distance function:
\begin{equation}
    \begin{aligned}
        &g_{\text{DI}}(x) = \\
        &\tanh \Big(\frac{(x_{1} - x_{1\text{,ref}})^2 + (x_{2}-x_{2\text{,ref}})^2 - r_{\text{DI target}}^2}{\alpha_{\text{DI}}}\Big),
    \end{aligned}
    \label{eq:g_double_integrator}
\end{equation}
where $r_{\text{DI target}}$ defines the target set radius and $\alpha_{\text{DI}}$ controls the sharpness of the transition between negative and positive values. The sign of $g(x)$ follows the calibration in \cref{ass:g-sign-calibration}, ensuring consistency with reachability semantics.

For the function approximation, we adopt a Sinusoidal Representation Network (SIREN) architecture, as prior work has shown that periodic activations are well-suited for representing value functions and their gradients in HJ settings \cite{bansal2021deepreach}.

\begin{table}[t]
\centering
\footnotesize
\caption{Parameters used in the double integrator reach experiment.}
\label{tab:exp-double-integrator}

\begin{tabularx}{\columnwidth}{@{}lX@{}}
\toprule
\textbf{Parameter} & \textbf{Value} \\
\midrule

Region of interest & $[-5,5]\times[-5,5]$ \\
Batch size $B$ & $16384$ \\
Learning rate & $10^{-4}$ \\
Optimizer & Adam \\
Training steps $N_{\text{train}}$ & $200{,}000$ \\

Sampling distribution $\rho$
&
$\frac{1}{2}\mathcal{U}([-5,5]\times[-5,5])$
$+\frac{1}{2}\mathcal{N}(x_{\mathrm{ref}},25I)$
\\

\midrule
\multicolumn{2}{c}{\textit{Environment}} \\
\midrule

Dynamics & Double integrator (\cref{eq:di_dyn}) \\
Control set $\mathcal{U}$ & $[-1,1]$ \\
Bellman step $\sigma$ (s) & $0.1$ \\
Time step $\Delta t$ (s) & $0.01$ \\
Rollout steps $H$ (-) & $10$ \\
Discount rate $\lambda$ (s$^{-1}$) & $1.0$ \\
Discount factor $\gamma$ (-) & $e^{-\lambda\sigma}\approx0.9512$ \\

\midrule
\multicolumn{2}{c}{\textit{Cost function}} \\
\midrule

Reach cost function & \cref{eq:g_double_integrator} \\
Target radius $r_{\text{target}}$ & $0.5$ \\
Smoothing parameter $\alpha$ (-) & $10.0$ \\

\midrule
\multicolumn{2}{c}{\textit{Neural network}} \\
\midrule

Network type & SIREN \\
Hidden layers & $(64,64)$ \\
Activation & $\sin(\cdot)$ \\
Activation frequency $\omega_0$ (rad/s) & $(10.0,1.0)$ \\

\bottomrule
\end{tabularx}
\end{table}

All parameters and hyperparameters used in the experiment associated with the double integrator are reported in \cref{tab:exp-double-integrator}.

\paragraph{Dubins Car Avoid Problem}

Furthermore, we consider a Dubins car avoid problem. The state of the Dubins car is described by the position $(x_1, x_2) \in \mathbb{R}^2$ and the heading angle $\theta \in [-\pi, \pi)$ rad. The control input $u$ represents the bounded turning rate while the forward velocity $\nu$ is held constant. The system evolves according to the following dynamics:
\begin{equation}
    \begin{aligned}
        \dot{x}_1 &= \nu \cos(\theta), \\
        \dot{x}_2 &= \nu \sin(\theta), \qquad u \in \mathcal{U}:=[-\dot{\theta}_{\text{max}}, \dot{\theta}_{\text{max}}] \\ 
        \dot{\theta} &= u(t),
    \end{aligned}   
    \label{eq:dyn-dubins-car}
\end{equation}

The goal is to avoid a certain region about a position ${x_{1\text{,ref}}, x_{2\text{,ref}}}$ and with radius $r_{\text{DC target}}$. Therefore, we design $g_{\text{DC}}(x)$ as:
\begin{equation}
    \begin{aligned}
        &g_{\text{DC}}(x) = \\
        &\tanh \Big(\frac{(x_{1} - x_{1\text{,ref}})^2 + (x_{2}-x_{2\text{,ref}})^2 - r_{\text{DC target}}^2}{\alpha_{\text{DC}}}\Big),
    \end{aligned}
    \label{eq:g_dubins_car_avoid}
\end{equation}

In contrast to the double integrator, the Dubins car avoid problem is expected to result in a non-trivial backward reachable tube. Consequently, it provides a meaningful test case for demonstrating that the proposed Bellman formulation preserves reachability semantics, namely that the discounting does not affect the resultant contour for the zero-level set.

\begin{table}[t]
\centering
\caption{Parameters used in the Dubins car avoid experiment. All unmentioned parameters are the same as in the case of the double integrator experiment (\cref{tab:exp-double-integrator}).}
\label{tab:exp-dubins-car}
\begin{tabular}{ll}
\toprule
\textbf{Parameter} & \textbf{Value} \\
\midrule

Region of interest & $[-5, 5] \times [-5, 5] \times [-\pi, \pi] $ \\
Learning rate & $2.87 \cdot 10^{-4}$ \\

\midrule
\multicolumn{2}{c}{\textit{Environment}} \\
\midrule

Dynamics & Dubins car (\cref{eq:dyn-dubins-car}) \\
$\nu$ (m/s) & 1.0 \\

\midrule
\multicolumn{2}{c}{\textit{Cost function}} \\
\midrule

Reach cost function & \cref{eq:g_dubins_car_avoid}\\

\midrule
\multicolumn{2}{c}{\textit{Neural network}} \\
\midrule

Hidden layers & $(128, 128)$\\
Activation frequency $\omega_0$ (rad/s) & $(25.9840, 1.2017)$ \\

\bottomrule
\end{tabular}
\end{table}

All parameters and hyperparameters used in the Dubins car avoid experiment are reported in \cref{tab:exp-dubins-car}.

\subsection{Validation of Bellman--HJ Equivalence}

We now focus on the discounted infinite-horizon setting and empirically verify that the proposed Bellman formulation recovers the same value function as the HJ characterization.

For both the double integrator and Dubins car, we compute:
\begin{itemize}
    \item a reference solution by solving the stationary HJ variational inequality (\cref{eq:HJVI-infinity}), and
    \item an approximation of the value function using FVI (\cref{alg:fvi-infinite}).
\end{itemize}

We examine pointwise differences between the value functions and the agreement of their zero level sets, which define the backward reachable tubes.

This provides empirical validation of \cref{thm:equivalence-characterization-infinity}, demonstrating that both the HJ and Bellman formulations yield the same value function. Furthermore, the close agreement supports the interpretation of reinforcement learning as a sample-based approximation of the Bellman fixed-point equation.

\begin{figure}[h!]
    \centering
    
    \begin{subfigure}[b]{\columnwidth}
        \centering
        \includegraphics[width=0.8\columnwidth]{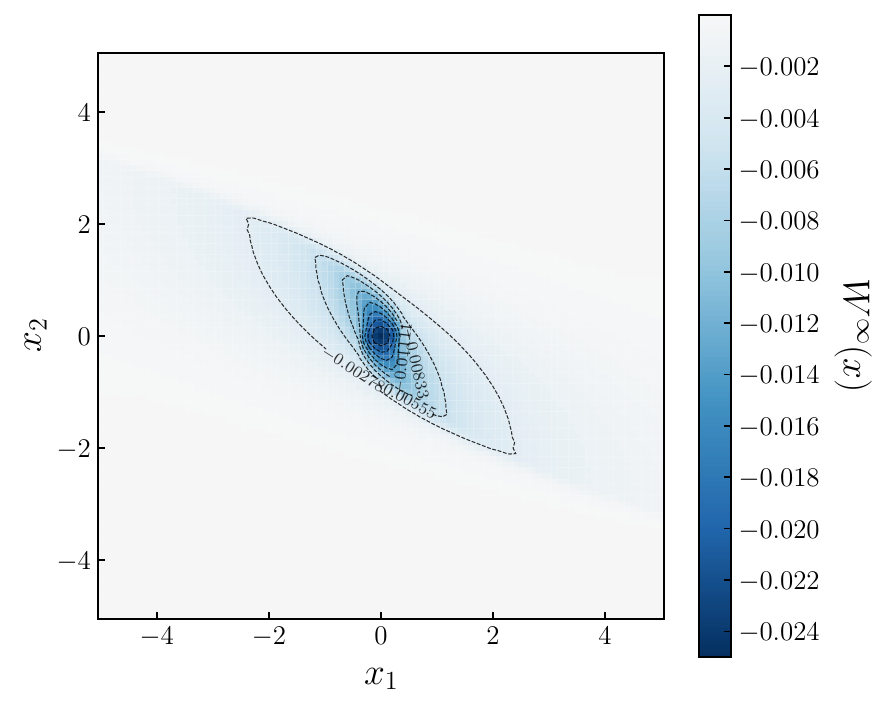}
        \caption{Reference value function obtained by solving stationary HJ variational inequality (\cref{eq:HJVI-infinity}) using semi-Lagrangian scheme.}
        \label{fig:ref-semilagrangian-double-integrator}
    \end{subfigure}
    
    \begin{subfigure}[b]{\columnwidth}
        \centering
        \includegraphics[width=0.8\columnwidth]{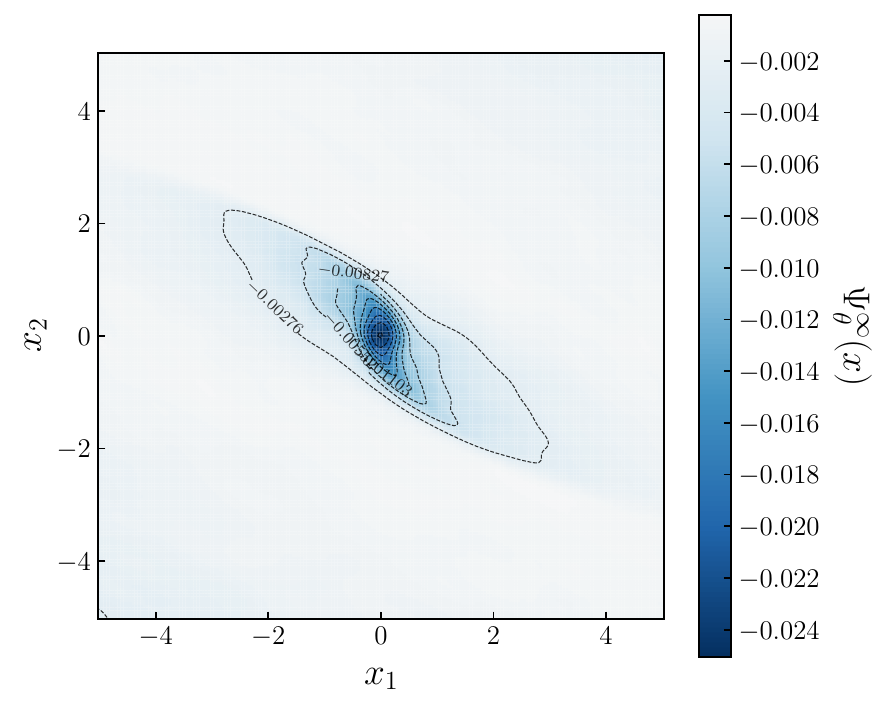}
        \caption{Value function learned using \cref{alg:fvi-infinite}.}
        \label{fig:ref-rl-double-integrator}
    \end{subfigure}

    \begin{subfigure}[b]{\columnwidth}
        \centering
        \includegraphics[width=0.8\columnwidth]{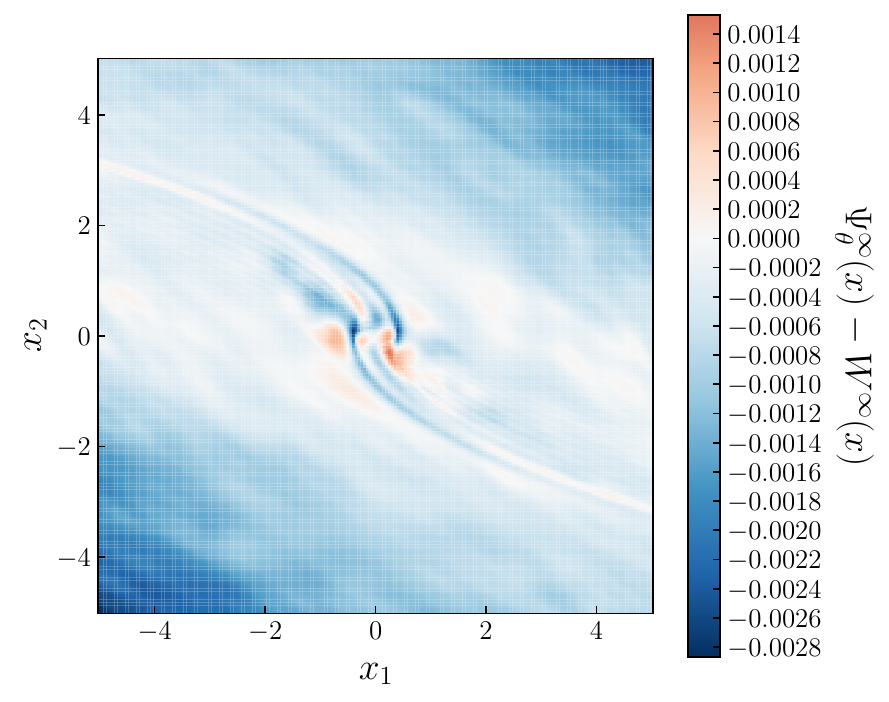}
        \caption{Pointwise difference between the learned and reference value functions.}
        \label{fig:diff-semilagrangian-rl-double-integrator}
    \end{subfigure}
    \caption{Comparison between the stationary solution of the HJ variational inequality (\cref{eq:HJVI-infinity}) and the value function learned via FVI, for the case of the double integrator reach problem.}
    \label{fig:experiment-double-integrator}
\end{figure}

Results are shown in \cref{fig:experiment-double-integrator}. The learned value functions closely match the reference solutions obtained from the semi-lagrangian scheme, with small pointwise error across the state space. These results support both the equivalence between the HJ and Bellman formulations and the interpretation of reinforcement learning as a sampled-based/data-driven solver of the Bellman fixed-point equation.

Moreover, as expected, since the double integrator is a controllable system, the reachable region becomes infinite when the horizon goes to infinity. That is why it is not possible to visualize the zero-level set which now spans the entire state space. In contrast, the Dubins car avoid problem yields a non-trivial backward reachable tube even in the infinite-horizon limit. Although the system is fully controllable, the avoid formulation induces a bounded safe set, defined as the set of states from which the system can avoid the unsafe region for all future times (see \cref{fig:experiment-dubins-car-avoid}). This provides a meaningful test case for evaluating whether the proposed Bellman formulation preserves reachability semantics (discussed in \cref{subsec:preservation-semantics}).

\subsection{Preservation of Reachability Semantics Under Discounting}
\label{subsec:preservation-semantics}

We empirically illustrate that discounting does not alter reachability semantics by comparing the zero level sets of the original undiscounted value function for the reach problem, namely,  
\begin{equation}
    \inf _{u(\cdot)} \min _{s \in[t, T]} g(x(s))
    \label{eq:og_value_function}
\end{equation}
and the zero-level set of the discounted value function \cref{eq:value_function}.

While the value functions themselves differ due to the presence of discounting, reachability is determined solely by the sign of the value function. Therefore, agreement of the signs and of the zero level sets indicates preservation of reachability semantics.

The comparison relies on a reference solution for the undiscounted value function which is obtained via the \texttt{helperOC} toolbox. Note that, in practice, the infinite-horizon solution cannot be obtained via \texttt{helperOC} as it requires a finite-time horizon. Thus, the infinite-horizon solution is approximated via a sufficiently long finite-time horizon.

\begin{figure}[h!]
    \centering

    \begin{subfigure}[b]{\columnwidth}
        \centering
        \includegraphics[width=0.8\columnwidth]{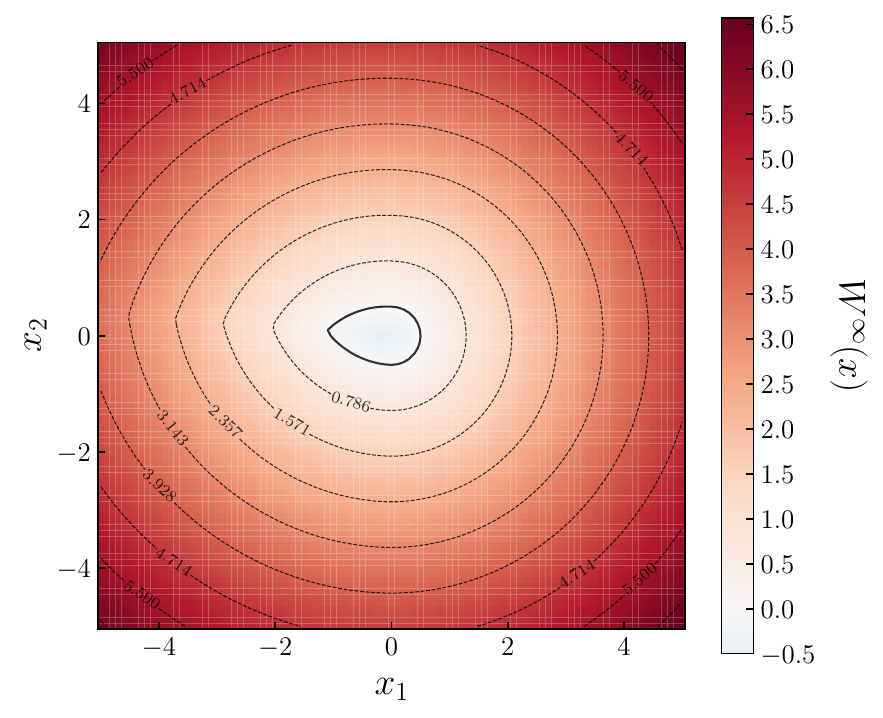}
        \caption{Reference value function for the undiscounted avoid problem, computed using \texttt{helperOC} over a long time horizon to approximate the infinite-horizon solution. The zero level set converges early and defines the BRT.}
        \label{fig:ref-helperOC-dubins-car}
    \end{subfigure}
    
    \begin{subfigure}[b]{\columnwidth}
        \centering
        \includegraphics[width=0.8\columnwidth]{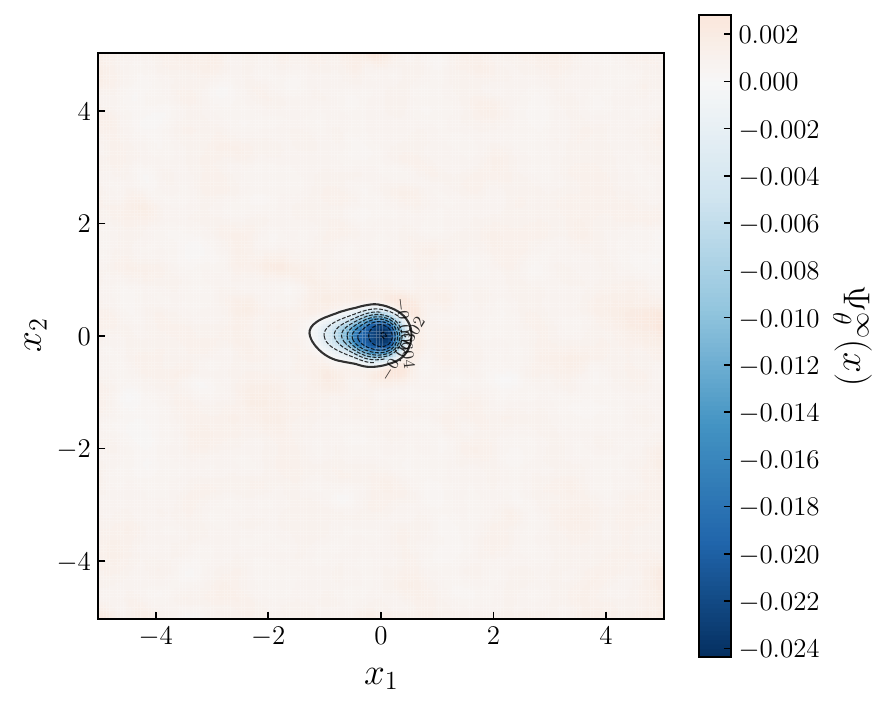}
        \caption{Value function learned using \cref{alg:fvi-infinite} as a sample-based approximation of the reachability-preserving Bellman operator.}
        \label{fig:ref-rl-dubins-car}
    \end{subfigure}
    
    \begin{subfigure}[b]{\columnwidth}
        \centering
        \includegraphics[width=0.65\columnwidth]{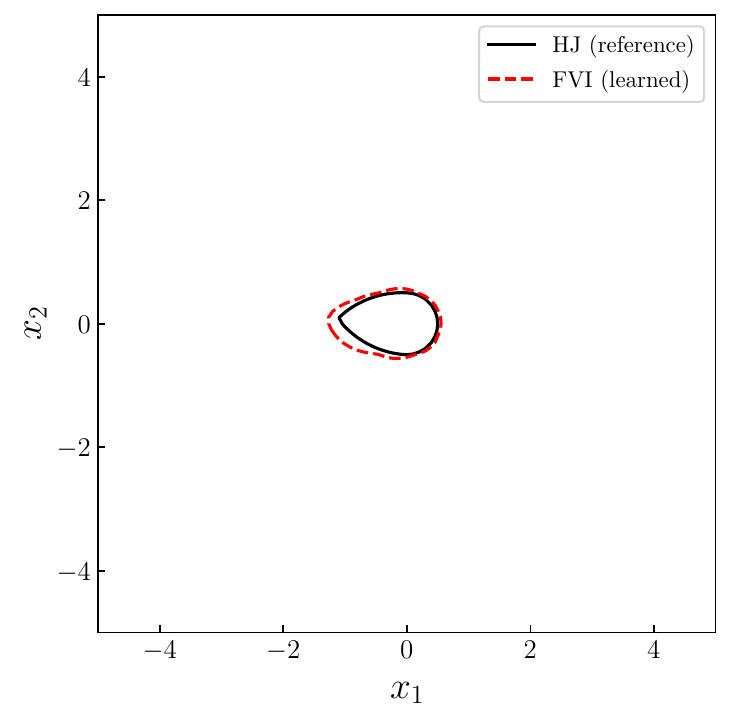}
        \caption{Comparison of zero level sets of the reference and learned value functions. Their alignment indicates preservation of reachability semantics under discounting.}
        \label{fig:contour-comparison-dubins-car}
    \end{subfigure}

    \caption{Dubins car avoid problem. Comparison between the undiscounted HJ reference solution and the value function learned via reinforcement learning. While the value magnitudes differ, the agreement of the zero level sets demonstrates preservation of reachability semantics.}

    \label{fig:experiment-dubins-car-avoid}
\end{figure}

The results are shown in \cref{fig:experiment-dubins-car-avoid}. The value functions exhibit different magnitudes, as expected, but their zero level sets are well aligned. This alignment indicates that discounting preserves the reachability semantics encoded by the value function.

\section{Conclusion}
\label{sec:conclusion}

This paper establishes a direct and exact connection between Hamilton–Jacobi reachability and reinforcement learning through a discounted reach-based value function that preserves reachability semantics. We showed that this value function admits two equivalent characterizations: as the unique viscosity solution of a Hamilton–Jacobi variational inequality and as the unique fixed point of a reachability-preserving Bellman operator.

A central result is that, despite its non-additive structure, the proposed Bellman operator is contractive under discounting. This yields a well-posed fixed-point problem with a unique solution and provides a principled foundation for computing reachability value functions via iterative methods.

Moreover, the equivalence established in this work provides a unifying perspective, and our results suggest that reinforcement learning can be interpreted as a sample-based method for approximating the same Bellman fixed point that characterizes the reachability value function. The results suggest a broader interpretation of reinforcement learning as a framework for approximating fixed points of suitable operators, rather than being solely tied to additive reward maximization. Importantly, this formulation preserves the exact semantics of the original reachability problem. This enables the use of learning-based methods in settings where formal guarantees are essential.

This work also shows that RL is not fundamentally tied to additive reward maximization, but can instead be understood as a method for approximating fixed points of more general operators. This enables the direct incorporation of reachability and safety semantics into learning, without relying on heuristic reward design.

Future work will focus on extending the framework to richer problem classes, including reach--avoid formulations and safety-critical control under disturbances, and stochastic dynamics. A further open direction concerns whether the sample-based approximation provably converges to the fixed point of the associated discrete Bellman operator during training (\cref{rem:fvi-convergence}). While \cref{sec:exp-res} provides empirical evidence of close agreement, an \emph{a priori} convergence guarantee for FVI on the non-additive reachability-preserving operator has not been established here. Independently of this question, the post-hoc residual certification framework of \cite{solanki2026certifying} may extend to the operators introduced here, which share the same underlying contraction property. Investigating this route further is also an interesting avenue for future work.

\ifieee
    \ifCLASSOPTIONcaptionsoff
      \newpage
    \fi
\fi

\bibliographystyle{plain}              
\bibliography{references}   

@article{solanki2026formalizing,
  title={Formalizing the Relationship between Hamilton-Jacobi Reachability and Reinforcement Learning},
  author={Solanki, Prashant and El-Hajj, Isabelle and van Beers, Jasper and van Kampen, Erik-Jan and de Visser, Coen},
  journal={arXiv preprint arXiv:2601.08050},
  year={2026}
}

@ARTICLE{akametalu2024minimum,
  author={Akametalu, Anayo K. and Ghosh, Shromona and Fisac, Jaime F. and Rubies-Royo, Vicenc and Tomlin, Claire J.},
  journal={IEEE Transactions on Automatic Control}, 
  title={A Minimum Discounted Reward Hamilton–Jacobi Formulation for Computing Reachable Sets}, 
  year={2024},
  volume={69},
  number={2},
  pages={1097-1103},
  doi={10.1109/TAC.2023.3327159}}

@book{bardi1997optimal,
  title={Optimal control and viscosity solutions of Hamilton-Jacobi-Bellman equations},
  author={Bardi, Martino and Dolcetta, Italo Capuzzo and others},
  volume={12},
  year={1997},
  publisher={Springer}
}

@book{granas2003fixed,
  title={Fixed point theory},
  author={Granas, Andrzej and Dugundji, James and others},
  volume={14},
  year={2003},
  publisher={Springer}
}

@inproceedings{fisac2019bridging,
  title={Bridging hamilton-jacobi safety analysis and reinforcement learning},
  author={Fisac, Jaime F and Lugovoy, Neil F and Rubies-Royo, Vicen{\c{c}} and Ghosh, Shromona and Tomlin, Claire J},
  booktitle={2019 International Conference on Robotics and Automation (ICRA)},
  pages={8550--8556},
  year={2019},
  organization={IEEE}
}

@inproceedings{bansal2021deepreach,
  title={Deepreach: A deep learning approach to high-dimensional reachability},
  author={Bansal, Somil and Tomlin, Claire J},
  booktitle={2021 IEEE International Conference on Robotics and Automation (ICRA)},
  pages={1817--1824},
  year={2021},
  organization={IEEE}
}

@article{sharpless2025dual,
  title={Dual-Objective Reinforcement Learning with Novel Hamilton-Jacobi-Bellman Formulations},
  author={Sharpless, William and Hirsch, Dylan and Tonkens, Sander and Shinde, Nikhil and Herbert, Sylvia},
  journal={arXiv preprint arXiv:2506.16016},
  year={2025}
}

@article{sharpless2026bellman,
  title={Bellman Value Decomposition for Task Logic in Safe Optimal Control},
  author={Sharpless, William and So, Oswin and Hirsch, Dylan and Herbert, Sylvia and Fan, Chuchu},
  journal={arXiv preprint arXiv:2602.19532},
  year={2026}
}

@article{hsu2021safety,
  title={Safety and liveness guarantees through reach-avoid reinforcement learning},
  author={Hsu, Kai-Chieh and Rubies-Royo, Vicen{\c{c}} and Tomlin, Claire J and Fisac, Jaime F},
  journal={arXiv preprint arXiv:2112.12288},
  year={2021}
}

@article{lee2020hopf,
  title={A Hopf-Lax formula in Hamilton--Jacobi analysis of reach-avoid problems},
  author={Lee, Donggun and Tomlin, Claire J},
  journal={IEEE Control Systems Letters},
  volume={5},
  number={3},
  pages={1055--1060},
  year={2020},
  publisher={IEEE}
}

@article{xue2019inner,
  title={Inner-approximating reachable sets for polynomial systems with time-varying uncertainties},
  author={Xue, Bai and Fr{\"a}nzle, Martin and Zhan, Naijun},
  journal={IEEE Transactions on Automatic Control},
  volume={65},
  number={4},
  pages={1468--1483},
  year={2019},
  publisher={IEEE}
}

@article{li2026converse,
  title={Converse barrier certificates for finite-time safety verification of continuous-time perturbed deterministic systems},
  author={Li, Yonghan and Wu, Chenyu and Wu, Taoran and Wang, Shijie and Xue, Bai},
  journal={Systems \& Control Letters},
  volume={209},
  pages={106357},
  year={2026},
  publisher={Elsevier}
}

@article{ganai2024hamilton,
  title={Hamilton-jacobi reachability in reinforcement learning: A survey},
  author={Ganai, Milan and Gao, Sicun and Herbert, Sylvia L},
  journal={IEEE Open Journal of Control Systems},
  volume={3},
  pages={310--324},
  year={2024},
  publisher={IEEE}
}

@mastersthesis{ganai2024hamiltonPhD,
  title={Hamilton-Jacobi Reachability Estimation in Reinforcement Learning},
  author={Ganai, Milan},
  year={2024},
  school={University of California, San Diego}
}

@inproceedings{bansal2017hamilton,
  title={Hamilton-jacobi reachability: A brief overview and recent advances},
  author={Bansal, Somil and Chen, Mo and Herbert, Sylvia and Tomlin, Claire J},
  booktitle={2017 IEEE 56th annual conference on decision and control (CDC)},
  pages={2242--2253},
  year={2017},
  organization={IEEE}
}

@inproceedings{herbert2021scalable,
  title={Scalable learning of safety guarantees for autonomous systems using hamilton-jacobi reachability},
  author={Herbert, Sylvia and Choi, Jason J and Sanjeev, Suvansh and Gibson, Marsalis and Sreenath, Koushil and Tomlin, Claire J},
  booktitle={2021 IEEE International Conference on Robotics and Automation (ICRA)},
  pages={5914--5920},
  year={2021},
  organization={IEEE}
}

@inproceedings{choi2021robust,
  title={Robust control barrier--value functions for safety-critical control},
  author={Choi, Jason J and Lee, Donggun and Sreenath, Koushil and Tomlin, Claire J and Herbert, Sylvia L},
  booktitle={2021 60th IEEE Conference on Decision and Control (CDC)},
  pages={6814--6821},
  year={2021},
  organization={IEEE}
}

@article{Mnih2015Human-level,
title={Human-level control through deep reinforcement learning},
author={Volodymyr Mnih and K. Kavukcuoglu and David Silver and Andrei A. Rusu and J. Veness and Marc G. Bellemare and Alex Graves and Martin A. Riedmiller and A. Fidjeland and Georg Ostrovski and Stig Petersen and Charlie Beattie and Amir Sadik and Ioannis Antonoglou and Helen King and D. Kumaran and Daan Wierstra and S. Legg and D. Hassabis},
journal={Nature},
year={2015},
volume={518},
pages={529-533},
doi={10.1038/nature14236}
}

@article{lygeros2004reachability,
  title={On reachability and minimum cost optimal control},
  author={Lygeros, John},
  journal={Automatica},
  volume={40},
  number={6},
  pages={917--927},
  year={2004},
  publisher={Elsevier}
}

@article{degrave2022magnetic,
  title={Magnetic control of tokamak plasmas through deep reinforcement learning},
  author={Degrave, Jonas and Felici, Federico and Buchli, Jonas and Neunert, Michael and Tracey, Brendan and Carpanese, Francesco and Ewalds, Timo and Hafner, Roland and Abdolmaleki, Abbas and de Las Casas, Diego and others},
  journal={Nature},
  volume={602},
  number={7897},
  pages={414--419},
  year={2022},
  publisher={Nature Publishing Group UK London}
}

@article{silver2016mastering,
  title={Mastering the game of Go with deep neural networks and tree search},
  author={Silver, David and Huang, Aja and Maddison, Chris J and Guez, Arthur and Sifre, Laurent and Van Den Driessche, George and Schrittwieser, Julian and Antonoglou, Ioannis and Panneershelvam, Veda and Lanctot, Marc and others},
  journal={nature},
  volume={529},
  number={7587},
  pages={484--489},
  year={2016},
  publisher={Nature Publishing Group UK London}
}

@article{silver2017mastering,
  title={Mastering the game of go without human knowledge},
  author={Silver, David and Schrittwieser, Julian and Simonyan, Karen and Antonoglou, Ioannis and Huang, Aja and Guez, Arthur and Hubert, Thomas and Baker, Lucas and Lai, Matthew and Bolton, Adrian and others},
  journal={nature},
  volume={550},
  number={7676},
  pages={354--359},
  year={2017},
  publisher={Nature Publishing Group UK London}
}

@article{silver2018general,
  title={A general reinforcement learning algorithm that masters chess, shogi, and Go through self-play},
  author={Silver, David and Hubert, Thomas and Schrittwieser, Julian and Antonoglou, Ioannis and Lai, Matthew and Guez, Arthur and Lanctot, Marc and Sifre, Laurent and Kumaran, Dharshan and Graepel, Thore and others},
  journal={Science},
  volume={362},
  number={6419},
  pages={1140--1144},
  year={2018},
  publisher={American Association for the Advancement of Science}
}

@article{vinyals2019grandmaster,
  title={Grandmaster level in StarCraft II using multi-agent reinforcement learning},
  author={Vinyals, Oriol and Babuschkin, Igor and Czarnecki, Wojciech M and Mathieu, Micha{\"e}l and Dudzik, Andrew and Chung, Junyoung and Choi, David H and Powell, Richard and Ewalds, Timo and Georgiev, Petko and others},
  journal={nature},
  volume={575},
  number={7782},
  pages={350--354},
  year={2019},
  publisher={Nature Publishing Group UK London}
}

@article{andrychowicz2020learning,
  title={Learning dexterous in-hand manipulation},
  author={Andrychowicz, OpenAI: Marcin and Baker, Bowen and Chociej, Maciek and Jozefowicz, Rafal and McGrew, Bob and Pachocki, Jakub and Petron, Arthur and Plappert, Matthias and Powell, Glenn and Ray, Alex and others},
  journal={The International Journal of Robotics Research},
  volume={39},
  number={1},
  pages={3--20},
  year={2020},
  publisher={SAGE Publications Sage UK: London, England}
}

@article{darbon2016algorithms,
  title={Algorithms for overcoming the curse of dimensionality for certain Hamilton--Jacobi equations arising in control theory and elsewhere},
  author={Darbon, J{\'e}r{\^o}me and Osher, Stanley},
  journal={Research in the Mathematical Sciences},
  volume={3},
  number={1},
  pages={19},
  year={2016},
  publisher={Springer}
}

@article{solanki2026certifying,
  title={Certifying Hamilton-Jacobi Reachability Learned via Reinforcement Learning},
  author={Solanki, Prashant and El-Hajj, Isabelle and van Beers, Jasper J. and van Kampen, Erik-Jan and de Visser, Coen C.},
  journal={arXiv preprint arXiv:2602.16475},
  year={2026},
  doi={10.48550/arXiv.2602.16475}
}

@book{bertsekas1996neuro,
  author    = {Bertsekas, Dimitri P. and Tsitsiklis, John N.},
  title     = {Neuro-Dynamic Programming},
  publisher = {Athena Scientific},
  address   = {Belmont, MA},
  year      = {1996}
}

@article{munos2005error,
  author  = {Munos, R{\'e}mi},
  title   = {Error Bounds for Approximate Value Iteration},
  journal = {Proceedings of the National Conference on Artificial Intelligence (AAAI)},
  year    = {2005}
}

@article{munos2008finite,
  author  = {Munos, R{\'e}mi and Szepesv{\'a}ri, Csaba},
  title   = {Finite-Time Bounds for Fitted Value Iteration},
  journal = {Journal of Machine Learning Research},
  volume  = {9},
  pages   = {815--857},
  year    = {2008}
}

\ifieee
\input{content/biographies}
\else
\fi

\end{document}